\documentclass[11pt,a4paper]{article}
\usepackage{amssymb}
\usepackage{multirow}
\usepackage{subfigure}
\usepackage[vlined,ruled]{algorithm2e}
\usepackage{color}
\usepackage{xcolor}
\usepackage{algpseudocode}
\SetAlFnt{\tiny}

\SetAlCapFnt{\normalfont\sffamily\tiny}

\makeatletter
\renewcommand{\algocf@Vline}[1]{%     no vskip in between boxes but a strut to separate them, 
  \strut\par\nointerlineskip% then interblock space stay the same whatever is inside it
  \algocf@push{\skiprule}%        move to the right before the vertical rule
  \hbox{\bgroup\color{cyan}\vrule\egroup%
    \vtop{\algocf@push{\skiptext}%move the right after the rule
      \vtop{\algocf@addskiptotal #1}\bgroup\color{cyan}\Hlne\egroup}}\vskip\skiphlne% inside the block
  \algocf@pop{\skiprule}%\algocf@subskiptotal% restore indentation
  \nointerlineskip}% no vskip after
\renewcommand{\algocf@Vsline}[1]{%    no vskip in between boxes but a strut to separate them, 
  \strut\par\nointerlineskip% then interblock space stay the same whatever is inside it
  \algocf@bblockcode%
  \algocf@push{\skiprule}%        move to the right before the vertical rule
  \hbox{\bgroup\color{cyan}\vrule\egroup%               the vertical rule
    \vtop{\algocf@push{\skiptext}%move the right after the rule
      \vtop{\algocf@addskiptotal #1}}}% inside the block
  \algocf@pop{\skiprule}% restore indentation
  \algocf@eblockcode%
}
\makeatother
\SetKwProg{Fn}{Function}{}{}

\usepackage{geometry}
\usepackage[compact]{titlesec}
\titlespacing{\section}{0pt}{*0}{*0}
\titlespacing{\subsection}{0pt}{*0}{*0}
\titlespacing{\subsubsection}{0pt}{*0}{*0}
\usepackage{graphicx}

\usepackage{times,latexsym}
\usepackage{url}
\usepackage[T1]{fontenc}
\usepackage{makecell}
\usepackage{authblk}
\usepackage[draft=false]{hyperref}
\usepackage[acceptedWithA]{tacl2018v2}

\usepackage{xspace,mfirstuc,tabulary}

\newif\iftaclinstructions
\taclinstructionsfalse % AUTHORS: do NOT set this to true
\iftaclinstructions
\renewcommand{\confidential}{}
\renewcommand{\anonsubtext}{(No author info supplied here, for consistency with
TACL-submission anonymization requirements)}
\newcommand{\instr}
\fi

\iftaclpubformat % this "if" is set by the choice of options

\else

\fi

\title{Fusing Visual, Textual and Connectivity Clues for Studying Mental Health}
\author[1]{Amir Hossein Yazdavar}
\author[1]{Mohammad Saeid Mahdavinejad}
\author[2]{Goonmeet Bajaj\\}
\author[3]{William Romine}
\author[1]{Amirhassan Monadjemi}
\author[1]{Krishnaprasad Thirunarayan\\}
\author[1]{Amit Sheth}
\author[4]{Jyotishman Pathak}
\affil[1]{Department of Computer Science \& Engineering, Wright State University, OH, USA}
\affil[2]{Ohio State University, Columbus, OH, USA}
\affil[3]{Department of  Biological Science, Wright State University, OH, USA}
\affil[4]{ Division of Health Informatics,  Weill Cornell University, New York, NY, USA}

\affil[1]{\textit {\ yazdavar.2@wright.edu}}
\date{}

\newcolumntype{H}{>{\setbox0=\hbox\bgroup}c<{\egroup}@{}}
\newcolumntype{Z}{>{\setbox0=\hbox\bgroup}c<{\egroup}@{\hspace*{-\tabcolsep}}}

\begin{document}

\setlength{\abovedisplayskip}{0pt}
\setlength{\belowdisplayskip}{0pt}
\maketitle
\begin{abstract}
  With ubiquity of social media platforms, millions of people are sharing their online persona by expressing their thoughts, moods, emotions, feelings, and even their daily struggles with mental health issues voluntarily and publicly on social media. Unlike the most existing efforts which study depression by analyzing textual content, we examine and exploit multimodal big data to discern depressive behavior using a wide variety of features including individual-level demographics. By developing a multimodal framework and employing statistical techniques for fusing heterogeneous sets of features obtained by processing visual, textual and user interaction data, we  significantly enhance the current state-of-the-art approaches for identifying depressed individuals on Twitter (improving the average F1-Score by 5 percent) as well as facilitate demographic inference from social media for broader applications. Besides providing insights into the relationship between demographics and mental health, our research assists in the design of a new breed of demographic-aware health interventions.
\end{abstract}

\iftaclpubformat

% Submission-specific rules

\fi

%\section{General instructions}
\section{Introduction}
Depression is a highly prevalent public health challenge and a major cause of disability worldwide.
Depression affects 6.7\% (i.e., about 16 million)  Americans  each year \footnote{\url{http://bit.ly/2okBKNy}}. According to the World Mental Health Survey conducted in 17 countries, on average, about 5\% of people reported having an episode of depression in 2011 \cite{marcus2012depression}.  Untreated or under-treated clinical depression can lead to suicide and other chronic risky behaviors such as drug or alcohol addiction\footnote{\url{https://wb.md/2pb4lm4}}. 
%More than 90\% of people who commit suicide already diagnosed with depression \cite{rudd2006warning}. 

%Apart from suicide, the tenth leading cause of death in the U.S. \cite{mathers2006projections}, depression has also other negative influences on daily, work, or school life, as well as sleeping and eating habits, and family or personal relationships \cite{de2013social}. Recent studies also show a strong association between mental disorder and chronic diseases such as cardiovascular diseases, diabetes, asthma \cite{kiang2015anxiety}, obesity and several adverse health behaviors like smoking \cite{bakhshaie2015cigarette}, physical inactivity and heavy drinking \cite{strine2008depression}.

Global efforts to curb clinical depression involve identifying depression through survey-based methods employing phone or online questionnaires. These approaches suffer from under-representation as well as sampling bias  (with very small group of respondents.)
%Besides, survey data also exhibit problems due to temporal gaps between the data collection and dissemination of findings, reflecting participant's responses over a short period of time. 
In contrast, the widespread adoption of social media where people voluntarily and publicly express their thoughts, moods, emotions, and feelings, and even share their daily struggles with mental health problems has not been adequately tapped into studying mental illnesses, such as depression. 
The visual and textual content shared on different social media platforms like Twitter offer new opportunities for a deeper understanding of self-expressed depression both at an individual as well as community-level. 
%For instance, the news headlines such as ``Twitter Fail:Teen Sent 144 Tweets Before Committing Suicide \& No One Helped'' and ``Study finds people still had limited access to mental healthcare from 2008-2014 as Affordable Care Act'' highlight the need for better tools for gleaning useful insights from user generated content on social media platforms that can assist policy designers in providing facilities for depressed individuals.
%Recent analysis have lead to  data-driven discoveriesalongside the typical hypothesis-testing social science process \cite{coppersmith2015clpsych,coppersmith2014measuring,andalibi2016understanding}.
Previous research efforts have suggested that language style, sentiment, users' activities, and engagement expressed in social media posts can predict the likelihood of depression \cite{de2013predicting,de2016discovering}. 
%These studies often use psycholinguistic analysis, supervised and unsupervised language modeling, and  expressed topics of interest.
%often gleaned by LIWC. 
However, except for a few attempts \cite{manikonda2017modeling,andalibi2017sensitive,reece2017instagram,ahsan2017towards}, these investigations have seldom studied extraction of emotional state from visual content of images in posted/profile images. Visual content can express users' emotions more vividly, and psychologists  noted that imagery is an effective medium for communicating difficult emotions\footnote{\url{https://bit.ly/2zpHQzw}}.
%Although, recently there are few attempts to study visual imagery disclosure of mental health on social media  (Instagram)\cite{manikonda2017modeling,andalibi2017sensitive} but we believe this modality of social media are yet unexplored.

%Indeed,  where visual imagery disclosure of mental health on social media  (Instagram)\cite{manikonda2017modeling,andalibi2017sensitive} has been taken into consideration, investigations in these fields are seldom concerned with language     expressions used to
%express such feelings.

According to eMarketer\footnote{\url{https://bit.ly/2rHtyGI}}, photos accounted for 75\% of content posted on Facebook worldwide and they are the most engaging type of content on Facebook (87\%).   Indeed,  "a picture is worth a thousand words" and now "photos are worth a million likes." Similarly, on Twitter, the tweets  with image links get twice as much attention as those without \footnote{\url{https://bit.ly/1u31GbO}}, and video-linked tweets drive up engagement \footnote{\url{https://bit.ly/2rQbKHj}}.  
The ease and naturalness of expression through visual imagery can serve to glean depression-indicators in vulnerable individuals who often seek social support through social media \cite{seabrook2016social}.   Further, as psychologist Carl Rogers highlights,  we often pursue and promote our Ideal-Self \footnote{\url{http://bit.ly/2hLnmqn}}. In this regard, the choice of profile image can be a proxy for the online persona \cite{liu2016analyzing}, providing a window into an individual's mental health status. For instance, choosing emaciated legs of girls covered with several cuts as profile image portrays negative self-view \cite{montesano2017depression}.

%or expressing emotions expression characterized by visual features
%and some people may prefer to chose pictures that do not stand for themselves 
%a teenager looks down at from a tall building while having suicidal thoughts 

%Moreover, psychologists argued that people use pictures to communicate messages[]
%social media posts in general These messages are often posted tend to represent The psychologist Carl Rogers provided a possible explanation for this, arguing that our personalities are composed of a “Real Self” (who we really are), and an “Ideal Self” (who we want to be).
%constantly motivated to pursue behaviors that bring us closer to our Ideal Self.
%However, despite some promising results in examining individual's depressive behavior from analyzing the textual generated content, ego-network, and image [] separately, to our knowledge, no research has been conducted to integrate all theses useful signals which help studying depression at the scale (community-level). 
%In particular, although psychologists argued that people use pictures to communicate messages[], few attempts have been conducted to deeply study the content of posted images[] or psychology of selecting profile picture[] and their association to depressive behavior. Furthermore,

 Inferring demographic information like gender and age can be crucial for stratifying our understanding of population-level epidemiology of mental health disorders. Relying  on  electronic health records data, previous studies explored gender differences in depressive behavior from different angles including prevalence, age at onset, comorbidities, as well as biological and psychosocial factors\footnote{\url{https://bit.ly/2P4kWs9}}. For instance,  women have been diagnosed with depression twice as often as men \cite{nolen1987sex} and national psychiatric morbidity survey in Britain has shown higher risk of depression in women \cite{mcmanus2016mental}. On the other hand, suicide rates for men are three to five times higher compared to that of the women \cite{angst2002gender}.
 %Women are more likely to socialize and express their dysphoria, 
 %while men tend to express their anger and show negative 
 %behavior's such as alcohol abuse and drug dependency \cite{meltzer1995prevalence}. 
 
Although depression can affect anyone at any age\footnote{\url{https://bit.ly/2JJeBia}}, signs and triggers of depression vary for different age groups \footnote{\url{https://bit.ly/2Rlyzzg}}. Depression triggers for children include parental depression, domestic violence, and loss of a pet, friend or family member. For teenagers (ages 12-18), depression may arise from hormonal imbalance, sexuality concerns and rejection by peers\footnote{\url{https://bit.ly/2qw2MRm}}. Young adults (ages 19-29) may develop depression due to life transitions, poverty, trauma, and work issues. Adult (ages 30-60) depression triggers include caring simultaneously for children and aging parents, financial burden, work and relationship issues. 
%Besides exhibiting common depression signs, these adults may succumb to drug and alcohol abuse, and other risky behaviors. 
Senior adults develop depression from common late-life issues, social isolation, major life loses such as the death of a spouse, financial stress and other chronic health problems (e.g., cardiac disease, dementia)\footnote{\url{https://wb.md/2D4jNJW}}. Therefore, inferring demographic information while studying depressive behavior from passively sensed social data, can shed better light on the population-level epidemiology of depression.     
%Late-life (above 65) depression is a common disease\footnote{https://www.webmd.com/mental-health/default.htm} that affects 6 million Americans. This causes suicide rate in people ages 80 to 84 to be more than twice that of the general population\footnote{https://www.webmd.com/depression/guide/depression-elderly}. Several studies highlight how depression impacts older people differently than younger people\cite{mirowsky1992age}. Depression in elderly occurs with other medical conditions often last longer, and increase risk of death\footnote{https://www.webmd.com/depression/guide/depression-elderly#1}. 
%Besides, even though there are some preliminary effort for connecting users location and depressive behavior(using twitter geo-tagged tweet[]), more accurately inferring users location will significantly enhance understanding the associations between socio-ecological and environmental phenomena (such as accessibility to mental health facilities[], economy[], environmental causes[]).

The recent advancements in deep neural networks, specifically for image analysis task, can 
lead to determining demographic features such as age and gender \cite{levi2015age}.
%Such crucial insights can assist policy designers in understanding associations between socio-ecological and environmental phenomena for effective community-level management of depression in various populations. 
 We show that by determining and integrating  heterogeneous set of features from different modalities --  aesthetic features from posted images (colorfulness, hue variance, sharpness, brightness, blurriness, naturalness), choice of profile picture (for gender, age, and facial expression), the screen name, the language features from both  textual content and profile's description (n-gram, emotion, sentiment), and finally sociability from ego-network, and user engagement -- we can reliably detect likely depressed individuals in a data set of 8,770 human-annotated Twitter users. 

We address and derive answers to the following research questions: 
1) How well do the content of posted images (colors, aesthetic and facial presentation) reflect depressive behavior?
2) Does the choice of profile picture show any psychological traits of depressed online persona? Are they reliable enough to represent the demographic information such as age and gender?
%and can they be used for community-level management of depression? 
3) Are there any underlying common themes among depressed individuals generated using multimodal content that can be used to detect depression reliably?

\begin{figure}[t!]
    \centering
    \includegraphics[width=0.53\textwidth]{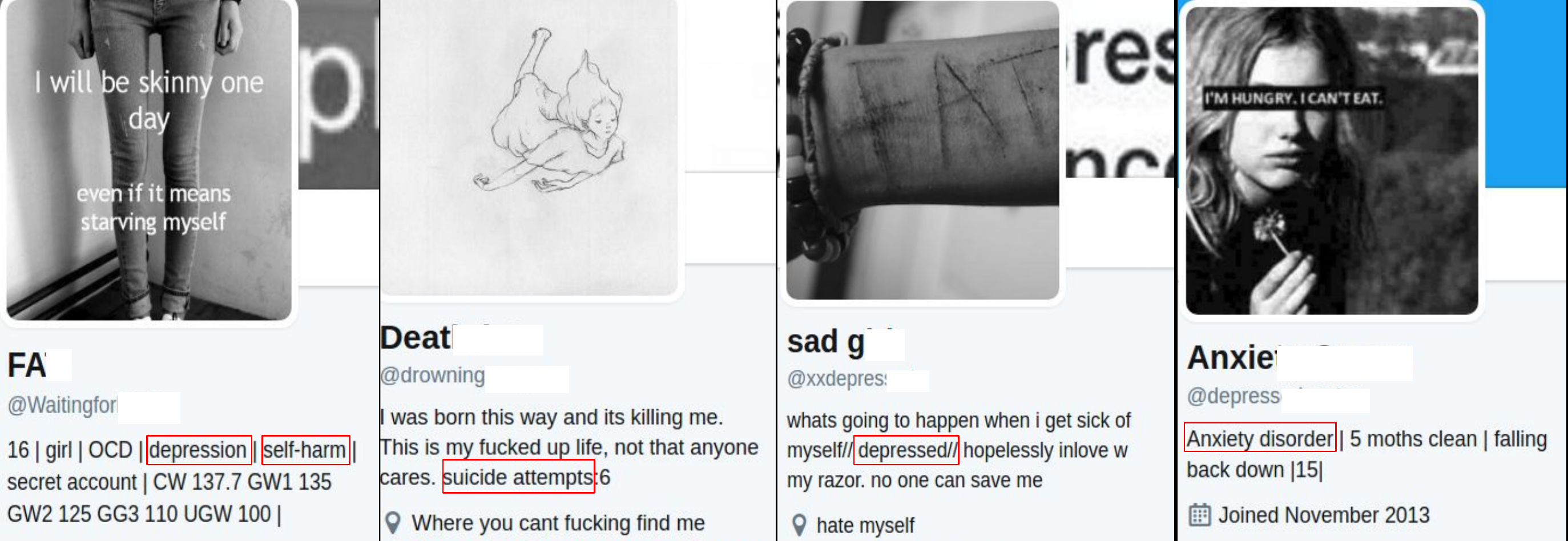}
    \setlength{\abovecaptionskip}{0pt plus 0pt minus 2pt}
    \vspace{-12pt}
    \caption{Self-disclosure on Twitter from likely depressed users discovered by matching depressive-indicative terms}
    \vspace{-8pt}
    \label{self_reported}
\end{figure}

%Diagnosing clinical depression requires consultation with mental health expert and clinicians. Our study aims to complement the traditional observational cohort studies conducted via online questionnaires by automatically detecting likely depressed individuals on Twitter. 

%\typeout{AMir-->TKP}: I tried to make them consistent. Does it help?

%\Taclpapers that do not comply with this document's instructions risk
%\iftaclpubformat
%publication delays until the camera-ready is brought into compliance.
%\else
%rejection without review.
%\fi

\section{Related Work}
%We divide the related work into two subsections. 
%We first discuss state-of-the-art approaches for studying depression on social data, then, we review studies on inferring demographic information.

\noindent
\textbf{Mental Health Analysis using Social Media:}
\noindent
Several efforts have attempted to automatically detect depression from social media content utilizing machine/deep learning and natural language processing approaches. Conducting a retrospective study over tweets, \cite{de2013social} characterizes depression based on factors such as language, emotion, style, ego-network, and user engagement. They built a classifier to predict the likelihood of depression in a post \cite{de2013social,shuai2016mining} or in an individual \cite{de2013predicting,nguyen2014affective, yazdavar2016analyzing, bajaj2017identifying}. Moreover, there have been significant advances due to the shared task \cite{coppersmith2015clpsych} focusing on methods for identifying depressed users on Twitter at the Computational Linguistics and Clinical Psychology Workshop (CLP 2015). A corpus of nearly 1,800 Twitter users was built for evaluation, and the best models employed topic modeling \cite{resnik2015beyond}, Linguistic Inquiry and Word Count (LIWC) features, and other metadata \cite{preotiuc2015role}. More recently, a neural network architecture introduced by \cite {yates2017depression} 
combined posts into a representation of
user's activities for detecting depressed users.
Another active line of research has focused on capturing suicide and self-harm signals \cite{coppersmith2018natural,thompson2014predicting,de2017language,wang2017understanding,de2016discovering,coppersmith2016exploratory}. Moreover, the CLP 2016 \cite{milne2016clpsych} defined a shared task on detecting the severity of the mental health from forum posts. \textit{All} of these studies derive discriminative features to classify depression in user-generated content at message-level, individual-level or community-level. Recent emergence of photo-sharing platforms such as Instagram, has attracted researchers attention to study people's behavior
from their visual narratives -- ranging from mining their emotions \cite{wang2015unsupervised}, and happiness trend \cite{abdullah2015collective}, to studying medical concerns \cite{garimella2016social}.  Researchers show that people use Instagram to engage in social exchange and storytelling about their difficult experiences \cite{andalibi2017sensitive}. The role of visual imagery as a mechanism of self-disclosure by relating visual attributes to mental health disclosures on Instagram was highlighted by \cite{manikonda2017modeling,reece2017instagram} where individual Instagram profiles were utilized to build a prediction framework for identifying markers of depression. 
The importance of data modality to understand user behavior on social media was highlighted by \cite{duong2017multimodal}.
More recently, a deep neural network sequence modeling approach that
marries audio and text data modalities to analyze question-answer style
interviews between an individual and an agent has been developed to study mental health \cite{duong2017multimodal}. Similarly, a multimodal depressive dictionary learning was proposed to detect depressed
users on Twitter \cite{shen2017depression}. They provide a sparse user representations by defining a feature set consisting of social network features, user profile features, visual features, emotional features \cite{ebrahimi2017challenges}, topic-level features, and domain-specific features.
Particularly, our choice of multi-model prediction framework is intended
to improve upon the prior works involving use of images in multimodal depression analysis \cite{shen2017depression} and prior works on studying Instagram photos \cite{ahsan2017towards,andalibi2016understanding}. 

\noindent
\textbf{Demographic information inference on Social Media: }
\noindent
There is a growing interest in understanding online user's demographic information due to its numerous applications in healthcare \cite{mislove2011understanding,lerman2016emotions}. A supervised model developed by \cite{burger2011discriminating} for determining users' gender by employing features such as screen-name, full-name, profile description and content on external resources (e.g., personal blog). Employing features including emoticons, acronyms, slangs, punctuations, capitalization, sentence length and included links/images, along with online behaviors such as number of friends,  post time, and commenting activity, a supervised model was built for predicting user's age group \cite{rosenthal2011age}. Utilizing users life stage information such as secondary school student, college student, and employee, \cite{nguyen2013old} builds age inference model for Dutch Twitter users.
Similarly, relying on profile descriptions while devising a set of rules and patterns, a novel model introduced for extracting age for Twitter users \cite{sloan2015tweets}. They also parse description for occupation by consulting the SOC2010 list of occupations\footnote{\url{https://www.bls.gov/soc/}} and validating it  through social surveys. 
A novel age inference model was developed while relying on homophily interaction information and content for predicting age of Twitter users \cite{zhang2016your}. %The intuition  is that people within the same age group share similar content and become  friends with contemporaries. 
%Using an extensive set of experiments, they show that their model outperformed other state-of-the-art age inference models by leveraging online interaction and content information simultaneously. 
The limitations of textual content for predicting age and gender was highlighted by \cite{nguyen2014gender}. They distinguish language use based on social gender,  age identity,  biological sex and chronological age by collecting crowdsourced signals using a game in which players (crowd) guess the biological sex and age of a user based  only on  their tweets. Their findings indicate how linguistic markers can misguide (e.g., a heart represented as <3 can be misinterpreted as feminine when the writer is male.) Estimating age and gender from facial images by training a convolutional neural networks (CNN) for face recognition is an active line of research \cite{han2013age,levi2015age,masi2016we}.

\section{Dataset}
%Self-disclosure refers to revealing personal and intimate information about oneself to others \cite{jourard1971self}. 
%It can be therapeutic for psychological well-being \cite{guntuku2017detecting}. Notably, individuals who share their unpleasant emotions enjoy greater well-being \cite{kennedy2001and}. 
%Previous efforts highlight diverse modes of mental health self-disclosures on social media \cite{manikonda2017modeling,andalibi2016understanding,yates2017depression}. 
Self-disclosure clues have been extensively utilized for creating ground-truth data for numerous social media analytic studies e.g., for predicting demographics \cite{mislove2011understanding,sloan2015tweets}, and user's depressive behavior \cite{yazdavar2017semi,de2017gender, 8419435}. For instance, vulnerable individuals may employ depressive-indicative terms in their Twitter profile descriptions. Others may share their age and gender, e.g., "16 years old suicidal girl"(see Figure \ref{self_reported}). We employ a huge dataset of 45,000 self-reported depressed users introduced in \cite{yazdavar2017semi} where a lexicon of depression symptoms consisting of 1500 depression-indicative terms was created with the help of psychologist clinician and employed for collecting self-declared depressed individual's profiles. A subset of 8,770 users (24 million time-stamped tweets) containing 3981 depressed and 4789 control users (that do not show any depressive behavior) were verified by two human judges \cite{yazdavar2017semi}. This dataset {$U_t$} contains the metadata \footnote{\url{https://bit.ly/2Wgsgke}} values of each user such as profile descriptions, followers\_count, created\_at, and profile\_image\_url. 

%We leverage these clues to build three different gold standard datasets.

%First, we selected a subset of highly informative depression-indicative terms from the depression lexicon  \cite{yazdavar2017semi} and found  matching profiles, obtaining ~45,000 Twitter users. Next,  we  generated a set of 5,000 random users.Note that these profiles also included  volunteer supporters, psychologists, psychiatrists,  medical experts, writers and filmmakers. Motivated by these observations, two human judges (native English speakers) were asked to manually verify if the collected accounts belonged to vulnerable individuals using PHQ-9 \footnote{PHQ-9 is a nine item depression scale clinicians use to screen, diagnose, and measure the severity of depression \url{http://www.cqaimh.org/pdf/tool_phq9.pdf}}.

%The average inter-annotator agreement was K=0.85 based on Cohen's Kappa statistics. After removing the profiles with less than 100 tweets and private accounts,  we obtained 8,770 users with 24 million time-stamped tweets, with each user contributing at most 3,200 tweets due to the Twitter Search API limitation. This dataset contained 3981 depressed and 4789 control users (that do not show any depressive behavior).  We denote this dataset of users by $U_t$.

%This dataset {$U_t$} contains the metadata \footnote{\url{https://developer.twitter.com/en/docs/tweets/data-dictionary/overview/user-object.html}} values of each user such as profile descriptions, followers\_count, created\_at, profile\_image\_url, etc. 

\begin{figure}[ht!]
    \centering
    \includegraphics[width=0.35\textwidth]{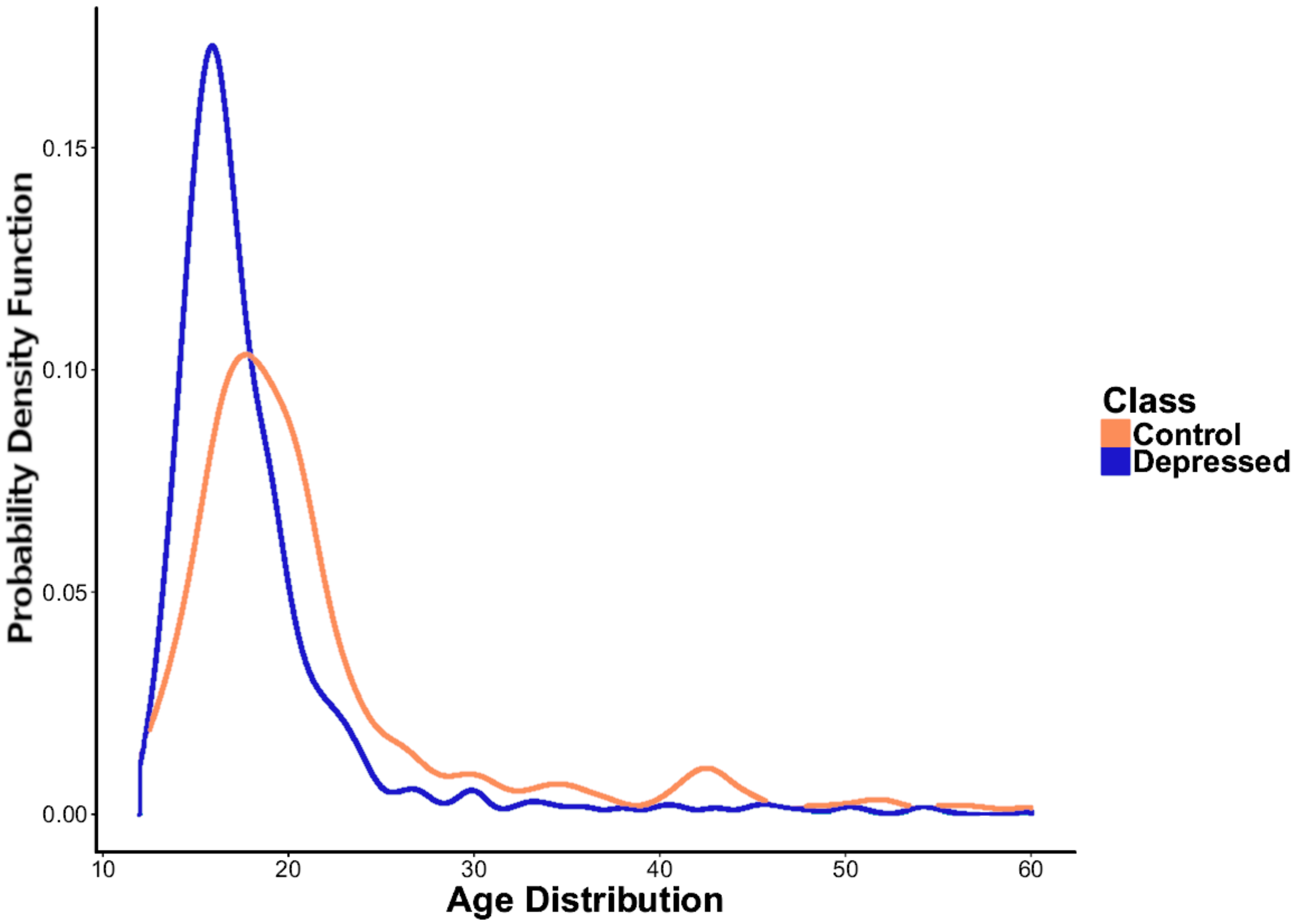}
    \setlength{\abovecaptionskip}{1pt plus 0pt minus 2pt}
    \caption{The age distribution for depressed and control users in ground-truth dataset}
    \vspace{-10pt}
    \label{age_dist}
\end{figure}

\noindent
\textbf{Age Enabled Ground-truth Dataset: }
We extract user's age by applying regular expression patterns to  profile descriptions (such as "17 years old, self-harm, anxiety, depression") \cite{sloan2015tweets}. 
%When a number is followed by the words "years" or "year," and is in age-range, we assume that a user is disclosing his/her age. Another  heuristic is when a unigram like "age" and/or bigrams such as "I am." preceded a number. 
We compile  "age prefixes" and  "age suffixes", and use three age-extraction rules: 1. I am X years old 2. Born in X 3. X years old, where X is a "date" or age (e.g., 1994). 
%Additionally, as a sanity check, we ask human annotator to validate the output of the obtained integers. Then, we adjust the obtained integer with "created\_at" metadata of user object that indicates the datetime a user account is created in Twitter. 
We selected a subset of 1061 users among $U_t$ as gold standard dataset $U_a$ who disclose their age. From these 1061 users, 822 belong to depressed class and 239 belong to control class. From 3981 depressed users, 20.6\% disclose their age in contrast with  only 4\% (239/4789) among control group. So self-disclosure of age is more prevalent among vulnerable users. Figure \ref{age_dist} depicts the age distribution in $U_a$. 
%The general trend, consistent with the results in \cite{zhang2016your,liao2014study}, shows that young people aged below 24 tend to be more depressed. 
The general trend, consistent with the results in \cite{zhang2016your,liao2014study}, is biased toward young people. Indeed, according to Pew, 47\% of Twitter users are younger than 30 years old \cite{duggan2015demographics}. Similar data collection procedure with comparable distribution have been
used in many prior efforts \cite{al2012homophily,liao2014study,zhang2016your}. We discuss our approach to mitigate the impact of the bias in Section 4.1. The median age is 17 for depressed class versus 19 for control class suggesting either likely depressed-user population is younger, or depressed youngsters are more likely to disclose their age for connecting to their peers (social homophily.) \cite{al2012homophily}

\begin{figure}[ht!]
    \centering
 \includegraphics[width=0.30\textwidth]{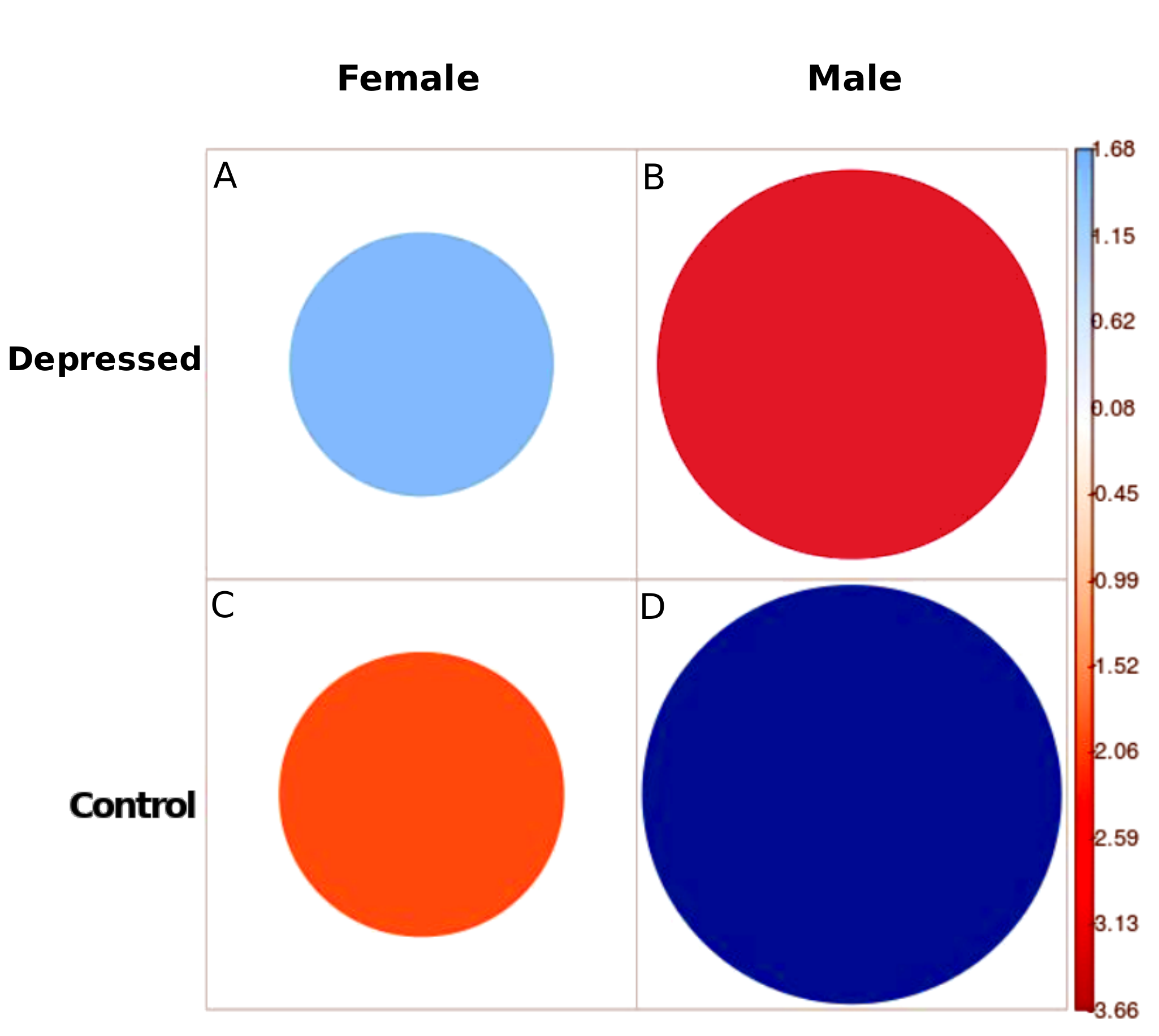}
 \setlength{\abovecaptionskip}{1pt plus 0pt minus 2pt}
    \caption{Gender and Depressive Behavior Association (Chi-square test: color-code: (blue:association), (red: repulsion), size: amount of each cell's contribution)}
    \vspace{-10pt}
    \label{gender_vs_depression}
\end{figure}

%The gold standard generate including only the users with complete interannotator agreement. 
\noindent
\textbf{Gender Enabled Ground-truth Dataset:} We selected a subset of 1464 users $U_g$ from $U_t$ who disclose their gender in their profile description. From 1464 users 64\% belonged to the depressed group, and the rest (36\%) to the control group. 23\% of the likely depressed users disclose their gender which is considerably higher (12\%) than that for the control class. Once again, gender disclosure varies among the two gender groups. For statistical significance, we performed chi-square test (null hypothesis: gender and depression are two independent variables). Figure \ref{gender_vs_depression} illustrates gender association with each of the two classes. Blue circles (positive residuals, see Figure \ref{gender_vs_depression}-A,D) show positive association among corresponding row and column variables while red circles (negative residuals, see Figure \ref{gender_vs_depression}-B,C) imply a repulsion. Our findings are consistent with the medical literature \cite{nolen1987sex} as according to \cite{ford2002prevalence} more women than men were given a diagnosis of depression. In particular, the female-to-male ratio is 2.1 and 1.9 for Major Depressive Disorder and Dysthymic Disorder respectively.  Our findings from Twitter data indicate there is a strong association (Chi-square: 32.75, p-value:1.04e-08) between being female and showing depressive behavior on Twitter. 

%Compile a lexicon of depression symptoms (background knowledge) that are likely to appear in the generated content of depressed individuals. 

%Urban self-reported, Gold standard, Twitter text Estimating age, gender, location from text(tool Washengton state) Image extraction correlation between feature from image and text, for 2 groups separately. personal profile table comparison between depressed vs non-depressed shared images analysis

\section{Data Modality Analysis}
We now provide an in-depth analysis of visual and textual content of vulnerable users.

\noindent
\textbf{Visual Content Analysis:}
We show that the visual content in images from posts as well as  profiles  provide valuable psychological cues  for understanding a user's depression status.  Profile/posted images can surface self-stigmatization \cite{barney2006stigma}. Additionally, as opposed to typical computer vision framework for object recognition that often relies on thousands of predetermined low-level features, what matters more for assessing user's online behavior is the  emotions reflected in facial expressions  \cite{pantic2009machine}, attributes contributing to the computational aesthetics \cite{datta2006studying}, and sentimental quotes they may subscribe to (Figure \ref{self_reported}) \cite{liu2016analyzing}.
%The following sections present an in-depth analysis of visual content for both the depressed and the control class with respect to three aspects -- facial presence, facial expressions and general image features.

\noindent
\textbf{Facial Presence: }
\noindent
 For capturing facial presence, we rely on \cite{zhou2013extensive}'s approach that uses multilevel convolutional coarse-to-fine 
network cascade  to tackle  facial landmark localization. We identify facial presentation,  emotion from facial expression, and demographic features from profile/posted images \footnote{\url{https://www.faceplusplus.com/}}.
Table \ref{facial-Presence} illustrates facial presentation differences in both profile and posted images (media) for depressed and control users in $U_t$. 
With control class showing significantly higher in both profile and media (8\%, 9\% respectively) compared to that for the depressed class. In contrast with age and gender disclosure,  vulnerable users are less likely to disclose their facial identity, possibly due to lack of confidence or fear of stigma.

\begin{table}
\tiny
\centering
\setlength{\abovecaptionskip}{1pt plus 0pt minus 2pt}
\caption{Facial Presence Comparison in Profile/Posted images for Depressed and Control Users}
\label{facial-Presence}
\begin{tabular}{cc|cc}
\hline
\multicolumn{1}{c|}{Face\_Found\_in} & \multicolumn{2}{c|}{\% Of Users}      & \ ${\chi}^2$ \\ \hline
                                       & Depressed & Control                   &                            \\ \hline
\multicolumn{1}{c|}{Media}             & 72\%      & \multicolumn{1}{c|}{81\%} & 163.52***                  \\
\multicolumn{1}{c|}{Profile}           & 4\%       & \multicolumn{1}{c|}{12\%} & 167.2***                   \\
\multicolumn{1}{c|}{Not\_found}      & 8\%       & \multicolumn{1}{c|}{7\%}  & 2.55                      
\end{tabular}
\vspace{-12pt}
\end{table}

\noindent \textbf{Facial Expression:}
\noindent
Following \cite{liu2016analyzing}'s approach, we adopt Ekman's model \footnote{\url{https://bit.ly/2TcNuO5}} of six emotions: anger, disgust, fear, joy, sadness and surprise, and use the Face++ API to automatically capture them from the shared images.  Positive emotions are joy and surprise, and negative emotions are anger, disgust, fear, and sadness. 
In general, for each user u in $U_t$, we process profile/shared images for both the depressed and the control groups with at least one face from the shared images (Table \ref{Statistics-image}). For the photos that contain multiple faces, we measure the average emotion.
\begin{table}
\tiny
\setlength{\abovecaptionskip}{1pt plus 0pt minus 2pt}
\caption{Statistics of Processed Shared/Profile  Images}
\label{Statistics-image}
\begin{tabular}{c|c|c|c}
\hline
\multicolumn{2}{c|}{\# of Proc. Prof. Images} & \multicolumn{2}{c}{\# of Proc. Shared Images} \\ \hline
Depressed                 & Control                 & \multicolumn{1}{c|}{Depressed}      & Control     \\ \hline
3466                      & 4127                    & \multicolumn{1}{c|}{265785}         & 401435     
\end{tabular}
\vspace{-10pt}
\end{table}
Figure \ref{fig:testEvaluation} illustrates the inter-correlation of these features. Additionally, we observe that emotions gleaned from facial expressions correlated with emotional signals captured from textual content utilizing LIWC. This indicates visual imagery can be harnessed as a complementary channel for measuring online emotional signals. 
%We will provide further details in following sections. 

\noindent
\textbf{General Image Features:}
\noindent
The importance of interpretable computational aesthetic features for studying users' online behavior has been highlighted by several efforts \cite{datta2006studying,liu2016analyzing,celli2014automatic}.  \textit{Color}, as a pillar of the human vision system, has a strong association with conceptual ideas like emotion \cite{naz2004relationship,huang2006natural}\footnote{\url{https://bit.ly/2DALcTq}}. We measured the normalized red, green, blue and the mean of original colors, and brightness and contrast  relative to variations of luminance. We represent images in \textit{Hue-Saturation-Value} color space that seems intuitive for humans, and measure mean and variance for saturation and hue. \textit{Saturation} is defined as the difference in the intensities of the different light wavelengths that compose the color. Although hue is not interpretable, high saturation indicates vividness and chromatic purity which are more appealing to the human eye \cite{liu2016analyzing}. 
\textit{Colorfulness} is measured as a difference against gray background \cite{san2009ranking}. \textit{Naturalness} is a measure of the degree of correspondence between images and the human perception of reality \cite{san2009ranking}. In color reproduction, \textit{naturalness} is measured from the mental recollection
of the colors of familiar objects. Additionally, there is a tendency among vulnerable users to share sentimental quotes bearing negative emotions. We performed optical character recognition (OCR) with python-tesseract \footnote{\url{https://pypi.org/project/pytesseract/}} to extract text and their sentiment score. As illustrated in Table \ref{t-test}, vulnerable users tend to use less colorful (higher grayscale) profile as well as  shared images  to convey their negative feelings, and share images that are less natural (Figure \ref{self_reported}). With respect to the aesthetic quality of images (saturation, brightness, and hue), 
%there is a significant difference as measured through their mean between the two classes, with 
depressed users use  images that are less appealing to the human eye. We employ independent t-test, while adopting Bonferroni Correction as a conservative approach to adjust the confidence intervals. Overall, we have 223 features, and choose Bonferroni-corrected $alpha$ level of $0.05/223= 2.24e-4$ (*** $p<alpha$, **$p<0.05$). 
%with a degree of freedom (df) of 740
%\subsection{Facial Expression}
\begin{table}
\centering
\tiny
\setlength{\abovecaptionskip}{1pt plus 0pt minus 2pt}
\caption[Caption for LOF]{Statistical significance (t-statistic) of the mean of salient features for depressed and control classes \footnotemark }
\footnotetext{Foot notes}
\label{t-test}
%\begin{tabular}{ll|c|c|c|c|c}
%\begin{tabular}{ll|c|c|c|c|c}
\begin{tabular}{p{3.5mm}p{14.5mm}|p{6.5mm}|p{4.5mm}|p{13mm}|p{8mm}H@{\hspace*{-\tabcolsep}}}
%{|p{0.111\textwidth}|p{0.312\textwidth}|}
\hline
\multicolumn{1}{c}{}                               & \multicolumn{1}{c|}{Feature} & \begin{tabular}[c]{@{}c@{}}Depressed\\ ($\mu$)\end{tabular} & \begin{tabular}[c]{@{}c@{}}Control\\ ($\mu$)\end{tabular} & \begin{tabular}[c]{@{}c@{}}95 per.\\ Conf.\\ interval\end{tabular} & T-stat  & P-value       \\ \hline
\multicolumn{1}{l|}{\multirow{12}{*}{Image-based}} & Prof.\_colorfulness        & 108                                                                   & 118.8                                                                 & (-15.38, -6.22)                                                     & -4.6***   & 4.25e-06***      \\
\multicolumn{1}{l|}{}                              & Prof.\_avgRGB         & 134.1                                                                     & 139                                                                   & ( 2.3 6.92)                                                    & -3.92*** & 9.38e-5***  \\
\multicolumn{1}{l|}{}                              & Prof.\_naturalness         & 0.3                                                                     & 0.6                                                                   & (-0.30, -0.19)                                                    & -12.7*** & 1.77e-32***  \\
\multicolumn{1}{l|}{}                              & Prof.\_hueVAR              & 0.05                                                                   & 0.07                                                                 & (-0.02, -0.008)                                                    & -4.6***   & 5.84e-6***     \\
\multicolumn{1}{l|}{}                              & Prof.\_Satu.VAR       & 0.03                                                                    & 0.04                                                                  & (-0.01, -0.003)                                                    & -3.9***   & 9.31e-5***       \\
\multicolumn{1}{l|}{}                              & Prof.\_Satu.Mean      & 0.2                                                                     & 0.31                                                                   & (-0.12, -0.07)                                                    & -8.9***   & 1.379e-11***  \\
\multicolumn{1}{l|}{}                              & Sha.\_BlueCh.Mean   & 119.5                                                                   & 134                                                             & (-9.82, -19.28)                                                     & -6***   & 3.20e-09***   \\
\multicolumn{1}{l|}{}                              & Sha.\_GraySc.Mean   & 0.5                                                                     & 0.49                                                                   & (0.03, 0.06)                                                       & 5.4***    & 1.87e-08***   \\
\multicolumn{1}{l|}{}                              & Sha.\_Colorfuln.    & 106.1                                                                   & 122                                                                 & (-14.9, -10.7)                                                    & -11.9***   & 2.55e-30***  \\
\multicolumn{1}{l|}{}                              & Sha.\_Satu.VAR   & 0.03                                                                    & 0.04                                                                  & (-0.01, -0.01)                                                     & -9.2*** & 1.24e-20***    \\
\multicolumn{1}{l|}{}                              & Sha.\_Satu.Mean  & 0.1                                                                    & 0.28                                                                  & (-0.10, -0.07)                                                    & -10.9***  & 6.17e-25***   \\
\multicolumn{1}{l|}{}                              & Sha.\_Naturalness     & 0.4                                                                    & 0.65                                                                  & (-0.19, -0.13)                                                    & -16.2***    & 2.0e-53***    \\ \hline
\multicolumn{1}{l|}{\multirow{6}{*}{Social-based}} & Friend.\_cnt               & 610.1                                                                  & 1380                                                                & (-1023,  -516)                                                      & -5.9***  & 5.9e-09***   \\
\multicolumn{1}{l|}{}                              & Followers\_cnt             & 589.4                                                                   & 1340                                                                & (-1148, -354)                                                    & -3.72**  & 0.0002**     \\
\multicolumn{1}{l|}{}                              & Stat.\_cnt              & 3722                                                                     & 7766                                                                   & (-6281, -1806)                                                      & -3.55**   & 0.0004**     \\
\multicolumn{1}{l|}{}                              & Avg\_fav.\_cnt  & 0.2                                                                     & 0.67                                                                   & (-0.78, -0.103)                                                    & -2.57   & 0.01*        \\
\multicolumn{1}{l|}{}                              & Avg\_retw.\_cnt   & 876.7                                                                   & 2720                                                                   & (-2673, -1013)                                                      & -4.3*** & 1.6e-05*** \\
\multicolumn{1}{l|}{}                              & Favourites\_cnt            & 2021                                                                     & 5199                                                               & (-5038, -1317)                                                      & -3.3**   & 8e-4**    
\end{tabular}
\vspace{-12pt}
\end{table}

\begin{figure*}[hptb]
\centering
\subfigure[\scriptsize{}]
{
\includegraphics[width=0.85\columnwidth]{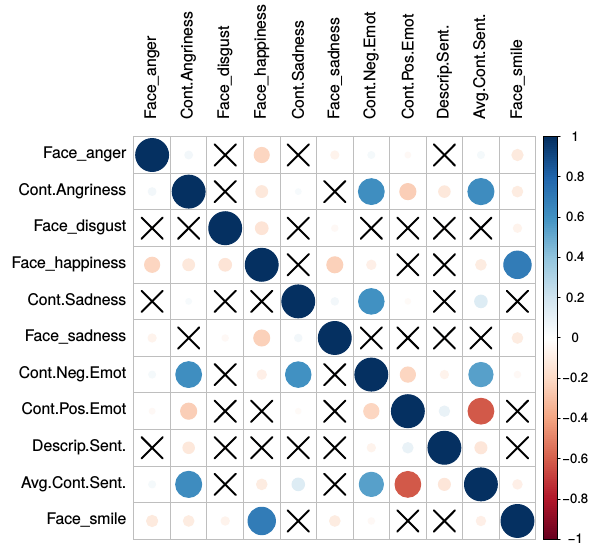}
\label{fig:QuerieswithMismatch}
}
~
\subfigure[\scriptsize {}]
%Pairs without statistically significant correlation are crossed (p-value <0.05)
{
\includegraphics[width=0.85\columnwidth]{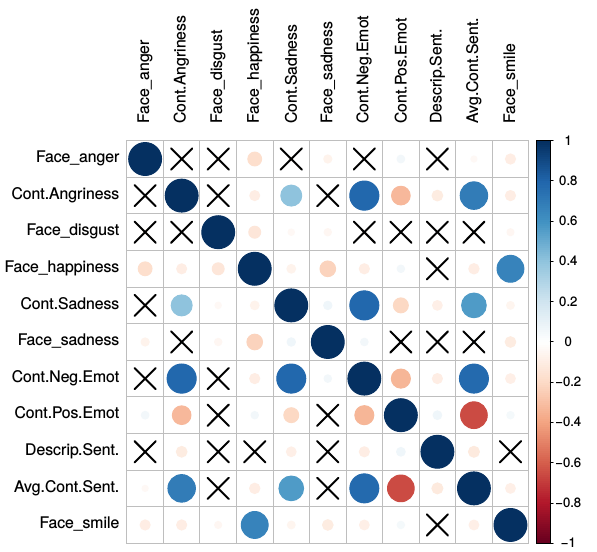}
\label{fig:QuerieswithNoMismatch}
}

%\caption{Correlation between Visual and Textual Content emotions}
\setlength{\abovecaptionskip}{1pt plus 0pt minus 2pt}
\vspace*{-3mm}
\caption{The Pearson correlation between the  average emotions derived from facial expressions through the shared images and emotions from textual content for \textit{depressed}-(a) and \textit{control users}-(b). Pairs without statistically significant correlation are crossed (p-value <0.05)}
\vspace{-15pt}
\label{fig:testEvaluation}
\end{figure*}

\typeout{Are you deriving emotion from both facial expression and content. Then we need to reword the caption further.----> yes,  any better??}

%------Footenote_17-------------
\footnotetext{** alpha= 0.05, *** alpha = 0.05/223}
%----------------------------

\noindent
\textbf{Demographics Inference \& Language Cues: }
 LIWC\footnote{\url{http://liwc.wpengine.com/}} 
%\typeout{footnote/supercript??}----->addressed
 has been used extensively for examining the latent dimensions of self-expression for analyzing personality \cite{schwartz2013personality}, depressive behavior, demographic differences \cite{nguyen2014gender,nguyen2013old}, etc. Several studies highlight that females employ more first-person singular pronouns \cite{chung2007psychological},  and deictic language\footnote{deictic: context-dependent words} \cite{mukherjee2010improving}, while males tend to use more articles \cite{argamon2007mining} which characterizes concrete thinking, and formal, informational and affirmation words \cite{newman2008gender}. For age analysis, the salient findings include older individuals using more future tense verbs \cite{chung2007psychological} triggering a shift in focus while aging. They also show positive emotions \cite{pennebaker2003words} and employ fewer self-references (i.e. 'I', 'me') with greater first person plural \cite{chung2007psychological}. Depressed users employ first person pronouns more frequently \cite{rude2004language}, repeatedly use negative emotions and anger words. We analyzed psycholinguistic cues and language style to study the association between depressive behavior as well as demographics.   
Particularly,  we adopt Levinson's adult development grouping \footnote{\url{https://bit.ly/2EuxUG8}} that partitions users in $U_a$ into 5 age groups: (14,19],(19,23], (23,34],(34,46], and (46,60]. 
Then, we apply LIWC for characterizing linguistic styles for each age group for users in $U_a$.

\textbf{Qualitative Language Analysis:} The recent LIWC version \footnote{\url{https://bit.ly/2PD8eQB}} summarizes textual content in terms of language variables such as analytical thinking, clout, authenticity, and emotional tone. It also measures other linguistic dimensions such as descriptors categories (e.g., percent of target words gleaned by dictionary, or longer than six letters - Sixltr) and informal language markers (e.g., swear words, netspeak), and other linguistic aspects (e.g., 1st person singular pronouns.)

\begin{figure}[ht!]
    \centering
    \includegraphics[width=0.50\textwidth]{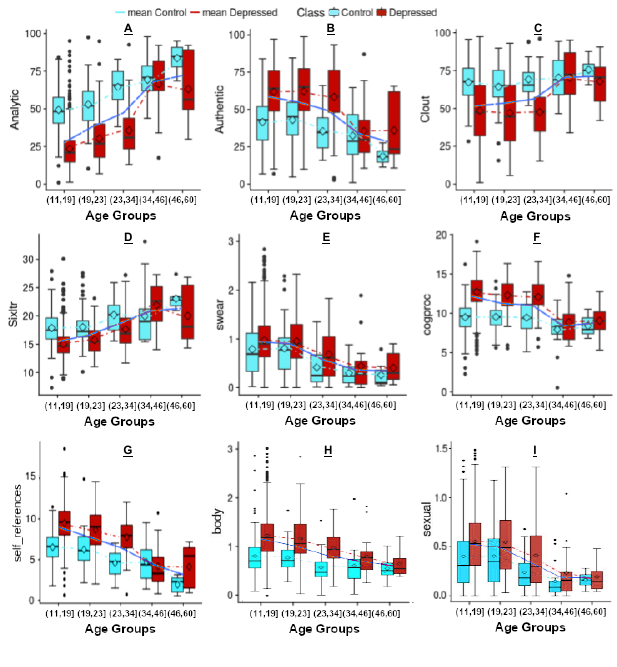}
    \setlength{\abovecaptionskip}{0pt plus 0pt minus 2pt}
    \vspace{-20pt}
    \caption{Characterizing Linguistic Patterns in two aspects: Depressive-behavior and Age Distribution}
    \vspace{-8pt}
    \label{LIWC-age}
\end{figure}

\noindent
\textbf{Thinking Style:}
\noindent
Measuring people's natural ways of trying to analyze, and organize complex events have strong association with analytical thinking. LIWC relates higher analytic thinking to more formal and logical reasoning whereas a lower value indicates focus on narratives. Also, cognitive processing measures problem solving in mind. Words such as "think," "realize," and "know" indicates the degree of "certainty" in communications. Critical thinking ability relates to education \cite{berger1984clinical}, and is impacted by different stages of cognitive development at different ages \footnote{\url{https://bit.ly/2znp77G}}. It has been shown that older people communicate with greater cognitive complexity while comprehending nuances and subtle differences \cite{chung2007psychological}. We observe a similar pattern in our data (Table \ref{Anova}.) A recent study highlights how depression affects brain and thinking at molecular level using a rat model \cite{calabrese2017chronic}. Depression can promote cognitive dysfunction including difficulty in concentrating and making decisions.
We observed a notable differences in the ability to think analytically in depressed and control users in different age groups (see Figure \ref{LIWC-age}- A, F and Table \ref{Anova}). Overall, vulnerable younger users are not logical thinkers based on their relative analytical score and cognitive processing ability.

\noindent
 \textbf{Authenticity:}
\noindent
 Authenticity measures the degree of honesty. Authenticity is often assessed by measuring present tense verbs, 1st person singular pronouns (I, me, my), and by examining the linguistic manifestations of false stories \cite{newman2003lying}. Liars
 use fewer self-references and fewer complex words. Psychologists often see a child's first successfull lie as a mental growth\footnote{\url{https://nyti.ms/2JDZlR7}}. There is a decreasing trend of the Authenticity with aging (see Figure \ref{LIWC-age}-B.)  
Authenticity for depressed youngsters is strikingly higher than their control peers. It decreases with age (Figure \ref{LIWC-age}-B.)  

\noindent 
\textbf{Clout:}
\noindent
People with high clout speak more confidently and with certainty, employing more social words with fewer negations (e.g., no, not) and swear words.
In general, midlife is relatively stable w.r.t. relationships and work. A recent study shows that age 60 to be best for self-esteem \cite{orth2018development} as people take on managerial roles at work and maintain a satisfying relationship with their spouse. We see the same pattern in our data (see Figure \ref{LIWC-age}-C and Table \ref{Anova}). Unsurprisingly, lack of confidence (the 6th PHQ-9 \footnote{https://bit.ly/2PY3INz} symptom) is a distinguishable characteristic of vulnerable users, leading to their lower clout scores,
especially among depressed users before middle age (34 years old). 

\noindent
%\textbf{Sixltr (Big words):} 
\noindent
%Employing big words (consisting of over six letters) which has strong association with higher grades and test scores (Figure \ref{LIWC-age}-D.)  
%Depressed users are more emotional\cite{park2012depressive} and avoid big words compared to control users.

%for each age group except for the 34-46 class where both the mean and the median for the depressed class is higher than control. 
\typeout{TKP->Amir: Your description of the connection between age, sixltr and emotion is confusing. Rewrite it clearly.---> addressed}

\noindent
\textbf{Self-references:}
\noindent
First person singular words are often seen as indicating interpersonal involvement and their high
usage is associated with negative affective states implying nervousness and depression \cite{pennebaker2003words}. Consistent with prior studies, frequency of first person singular for depressed people is significantly higher compared to that of control class. Similarly to \cite{pennebaker2003words}, youngsters tend to use more first-person (e.g. I) and second person singular (e.g. you) pronouns (Figure \ref{LIWC-age}-G). 
%The impact of the above phenomenon reflected in significantly higher frequency of self-references for depressed youngsters.  As with control class, a downtrend suggests as depressed individual age, they make more distinctions and psychologically distance themselves from their topics. 

%THey tend to be more honest
\noindent
\textbf{Informal Language Markers; Swear, Netspeak:}
\noindent
%Swear lexicon includes terms such as fu**, dam*, and shi*.
Several studies highlighted the use of profanity by young adults has significantly increased over the last decade \cite{kaye2004watch}. 
We observed the same pattern in both the depressed and the control classes (Table \ref{Anova}), although it's rate is higher for depressed users \cite{de2013predicting}. Psychologists have also shown that swearing can indicate that an individual is not a fragmented member of a society\footnote{\url{https://bit.ly/2RRqV4U}}. Depressed youngsters, showing higher rate of interpersonal involvement and relationships, have a higher rate of cursing (Figure \ref{LIWC-age}-E).  Also, Netspeak lexicon measures the frequency of terms such as lol and thx. 
%Although the rate is higher for depressed class, we did not find any pattern concerning adult development. 

\noindent
\textbf{Sexual, Body: } 
\noindent
Sexual lexicon contains terms like "horny", "love" and "incest", and body terms like "ache", "heart", and "cough". Both start with a higher rate for depressed users while decreasing gradually while growing up, possibly due to changes in sexual desire as we age \footnote{\url{https://bit.ly/2RyrcF5}} (Figure \ref{LIWC-age}-H,I and Table \ref{Anova}.)

%\begin{figure}[!t]
    %\centering
    %\includegraphics[width=0.53\textwidth]{figs/Liwc_age_f}
    %\caption{Characterizing latent dimensions of self-disclosure in two aspects: depressive-behavior and age distribution}
 %   \label{fig_trends}
%\end{figure}

%\begin{figure}[!t]
 %   \centering
  %  \includegraphics[width=0.53\textwidth]{figs/age_analysis_2}
   % \caption{Age distribution for the both group of Depressed and Control}
%    \label{fig_trends}
%\end{figure}

%\begin{figure}[!t]
 %   \centering
  %  \includegraphics[width=0.53\textwidth]{figs/age_analysis3}
   % \caption{Age distribution for the both group of Depressed and Control}
   % \label{fig_trends}
%\end{figure}
%\noindent
\textbf{Quantitative Language Analysis:}
\noindent
 We employ one-way ANOVA to compare the impact of various factors and validate our findings above. Table \ref{Anova} illustrates our findings, with a degree of freedom (df) of 1055. The null hypothesis is that the sample means' for each age group are similar for each of the LIWC features.

%*** alpha = 0.001, ** alpha = 0.01,  * alpha = 0.05 
% Please add the following required packages to your document preamble:
% \usepackage{multirow}
\begin{table}
\tiny
\centering
%\caption[Caption for LOF]{Statistical significance test using one-way ANOVA\textsuperscript{a=}}
%\small\textsuperscript{a=} The footnote-like comment under the caption
%\caption{Statistical significance test using one-way ANOVA %\protect\footnote{*** alpha = 0.001, ** alpha = 0.01,  * alpha = 0.05} }
\setlength{\abovecaptionskip}{1pt plus 0pt minus 2pt}
\caption[Caption for LOF]{Statistical Significance Test of Linguistic Patterns/Visual Attributes for Different Age Groups with one-way ANOVA \footnotemark }
\label{Anova}
\begin{tabular}{p{1mm}|l|p{3.7mm}|p{3.7mm}|p{3.7mm}|p{3.7mm}|p{3.7mm}|p{0.1mm}}
%\begin{tabular}{c|c|l|l|l|l|l|l|l}
\hline
\multirow{13}{*}{Text-based}                      & \multirow{2}{*}{Feature}                                   & \multicolumn{5}{c|}{\begin{tabular}[c]{@{}c@{}}Mean\\ (SD)\end{tabular}}                                                                                                                                                                                                                             & \multicolumn{1}{c}{\multirow{2}{*}{F-value}} \\ \cline{3-7}
                                                  &                                                            & {[}11,19)                                                & {[}19,23)                                                & {[}23,34)                                                & {[}34,46)                                                & {[}46,60)                                                & \multicolumn{1}{c}{}                         \\ \cline{2-8} 
                                                  & Analytic                                                   & \begin{tabular}[c]{@{}l@{}}27.62\\ (16.62)\end{tabular}  & \begin{tabular}[c]{@{}l@{}}38.61\\ (19.16)\end{tabular}  & \begin{tabular}[c]{@{}l@{}}47.28\\ (20.69)\end{tabular}  & \begin{tabular}[c]{@{}l@{}}67.88\\ (18.51)\end{tabular}  & \begin{tabular}[c]{@{}l@{}}72.05\\ (20.79)\end{tabular}  & 84***                                     \\
                                                  & Authentic                                                  & \begin{tabular}[c]{@{}l@{}}58.54\\ (19.54)\end{tabular}  & \begin{tabular}[c]{@{}l@{}}55.04\\ (20.04)\end{tabular}  & \begin{tabular}[c]{@{}l@{}}49.21\\ (22.05)\end{tabular}  & \begin{tabular}[c]{@{}l@{}}33.99\\ (19.73)\end{tabular}  & \begin{tabular}[c]{@{}l@{}}28.39\\ (19.04)\end{tabular}  & 22***                                     \\
                                                  & Clout                                                      & \begin{tabular}[c]{@{}l@{}}51.6\\ (21.35)\end{tabular}   & \begin{tabular}[c]{@{}l@{}}53.43\\ (21.26)\end{tabular}  & \begin{tabular}[c]{@{}l@{}}56.27\\ (19.81)\end{tabular}  & \begin{tabular}[c]{@{}l@{}}70.28\\ (17.46)\end{tabular}  & \begin{tabular}[c]{@{}l@{}}71.21\\ (13.50)\end{tabular}  & 9***                                       \\
                                                  & Dic                                                        & \begin{tabular}[c]{@{}l@{}}85.04 \\ (6.06)\end{tabular}  & \begin{tabular}[c]{@{}l@{}}82.63 \\ (6.21)\end{tabular}  & \begin{tabular}[c]{@{}l@{}}80.48\\ (6.56)\end{tabular}   & \begin{tabular}[c]{@{}l@{}}75.87\\ (6.91)\end{tabular}   & \begin{tabular}[c]{@{}l@{}}74.09\\ (5.95)\end{tabular}   & 37***                                     \\
                                                  & Article                                                    & \begin{tabular}[c]{@{}l@{}}3.52\\ (0.78)\end{tabular}    & \begin{tabular}[c]{@{}l@{}}3.92\\ (0.73)\end{tabular}    & \begin{tabular}[c]{@{}l@{}}4.00\\ (0.80)\end{tabular}    & \begin{tabular}[c]{@{}l@{}}4.52\\ (1.38)\end{tabular}    & \begin{tabular}[c]{@{}l@{}}5.13\\ (1.00)\end{tabular}    & 35***                                     \\
                                                  & Sixltr                                                     & \begin{tabular}[c]{@{}l@{}}15.48\\ (2.84)\end{tabular}   & \begin{tabular}[c]{@{}l@{}}16.58\\ (3.07)\end{tabular}   & \begin{tabular}[c]{@{}l@{}}18.65\\ (3.71)\end{tabular}   & \begin{tabular}[c]{@{}l@{}}20.88\\ (4.74)\end{tabular}   & \begin{tabular}[c]{@{}l@{}}21.33\\ (4.11)\end{tabular}   & 52***                                     \\
                                                  & \begin{tabular}[c]{@{}c@{}}Cogn. words\end{tabular} & \begin{tabular}[c]{@{}l@{}}12.17\\ (2.53)\end{tabular}   & \begin{tabular}[c]{@{}l@{}}11.24\\ (2.38)\end{tabular}   & \begin{tabular}[c]{@{}l@{}}10.99\\ (2.55)\end{tabular}   & \begin{tabular}[c]{@{}l@{}}8.36\\ (2.63)\end{tabular}    & \begin{tabular}[c]{@{}l@{}}8.75\\ (1.96)\end{tabular}    & 28***                                     \\
                                                  & Self-ref                           & \begin{tabular}[c]{@{}l@{}}14.13\\ (2.35)\end{tabular}   & \begin{tabular}[c]{@{}l@{}}12.45\\ (2.56)\end{tabular}   & \begin{tabular}[c]{@{}l@{}}10.96\\ (2.60)\end{tabular}   & \begin{tabular}[c]{@{}l@{}}9.05\\ (3.69)\end{tabular}    & \begin{tabular}[c]{@{}l@{}}7.55\\ (3.38)\end{tabular}    & 85***                                     \\
                                                  & Swear                                                      & \begin{tabular}[c]{@{}l@{}}0.96\\ (0.59)\end{tabular}    & \begin{tabular}[c]{@{}l@{}}0.89\\ (0.53)\end{tabular}    & \begin{tabular}[c]{@{}l@{}}0.57\\ (0.48)\end{tabular}    & \begin{tabular}[c]{@{}l@{}}0.36\\ (0.41)\end{tabular}    & \begin{tabular}[c]{@{}l@{}}0.33\\ (0.30)\end{tabular}    & 18***                                     \\
                                                  & Money                                                      & \begin{tabular}[c]{@{}l@{}}0.27\\ (0.40)\end{tabular}    & \begin{tabular}[c]{@{}l@{}}0.38\\ (0.19)\end{tabular}    & \begin{tabular}[c]{@{}l@{}}0.45\\ (0.25)\end{tabular}    & \begin{tabular}[c]{@{}l@{}}0.52\\ (0.22)\end{tabular}    & \begin{tabular}[c]{@{}l@{}}0.78\\ (0.37)\end{tabular}    & 15***                                     \\
                                                  & Work                                                       & \begin{tabular}[c]{@{}l@{}}0.80\\ (0.39)\end{tabular}    & \begin{tabular}[c]{@{}l@{}}1.09\\ (0.53)\end{tabular}    & \begin{tabular}[c]{@{}l@{}}1.31\\ (0.76)\end{tabular}    & \begin{tabular}[c]{@{}l@{}}1.67\\ (0.83)\end{tabular}    & \begin{tabular}[c]{@{}l@{}}2.02\\ (1.01)\end{tabular}    & 69***                                     \\ \hline
\multicolumn{1}{l|}{\multirow{5}{*}{Image-based}} & \multicolumn{1}{l|}{Prof.\_natu.}                    & \begin{tabular}[c]{@{}l@{}}37.80\\ (13.84)\end{tabular}  & \begin{tabular}[c]{@{}l@{}}48.05\\ (18.64)\end{tabular}  & \begin{tabular}[c]{@{}l@{}}52.33\\ (28.51)\end{tabular}  & \begin{tabular}[c]{@{}l@{}}64.33\\ (24.53)\end{tabular}  & \begin{tabular}[c]{@{}l@{}}68.07\\ (15.28)\end{tabular}  & 10***                                     \\
\multicolumn{1}{l|}{}                             & \multicolumn{1}{l|}{Prof.\_Satu.Mean}                  & \begin{tabular}[c]{@{}l@{}}20.31\\ (1.95)\end{tabular}   & \begin{tabular}[c]{@{}l@{}}23.27\\ (1.99)\end{tabular}   & \begin{tabular}[c]{@{}l@{}}29.78\\ (1.99)\end{tabular}   & \begin{tabular}[c]{@{}l@{}}38.76\\ (2.14)\end{tabular}   & \begin{tabular}[c]{@{}l@{}}33.13\\ (1.94)\end{tabular}   & 9***                                      \\
\multicolumn{1}{l|}{}                             & \multicolumn{1}{l|}{Prof.\_Colorful.}                 & \begin{tabular}[c]{@{}l@{}}106.47\\ (42.70)\end{tabular} & \begin{tabular}[c]{@{}l@{}}107.95\\ (39.15)\end{tabular} & \begin{tabular}[c]{@{}l@{}}111.01\\ (42.09)\end{tabular} & \begin{tabular}[c]{@{}l@{}}113.97\\ (35.48)\end{tabular} & \begin{tabular}[c]{@{}l@{}}123.60\\ (27.60)\end{tabular} & 0.89                                         \\
\multicolumn{1}{l|}{}                             & \multicolumn{1}{l|}{Shared\_avgRGB}                         & \begin{tabular}[c]{@{}l@{}}139.20\\ (18.12)\end{tabular} & \begin{tabular}[c]{@{}l@{}}140.45\\ (16.00)\end{tabular} & \begin{tabular}[c]{@{}l@{}}131.55\\ (16.32)\end{tabular} & \begin{tabular}[c]{@{}l@{}}133.74\\ (22.41)\end{tabular} & \begin{tabular}[c]{@{}l@{}}139.02\\ (22.30)\end{tabular} & 3**                                       \\
\multicolumn{1}{l|}{}                             & \multicolumn{1}{l|}{Prof.\_GrayMean}                   & \begin{tabular}[c]{@{}l@{}}0.471\\ (0.19)\end{tabular}   & \begin{tabular}[c]{@{}l@{}}0.474\\ (0.16)\end{tabular}   & \begin{tabular}[c]{@{}l@{}}0.456\\ (0.21)\end{tabular}   & \begin{tabular}[c]{@{}l@{}}0.470\\ (0.14)\end{tabular}   & \begin{tabular}[c]{@{}l@{}}0.450\\ (0.11)\end{tabular}   & 0.12                                         \\ \hline
\end{tabular}
\vspace{-14pt}
\end{table}

%------Footenote-------------
\footnotetext{*** alpha = 0.001, ** alpha = 0.01,  * alpha = 0.05 }
%----------------------------

\subsection{Demographic Prediction}
%Online behavior is representative of both a user's demographics and mental health status.
We leverage both the visual and textual content for predicting age and gender. 

\noindent
%\subsection{Gender Prediction}
\textbf{Prediction with Textual Content:}
\noindent
We employ \cite{sap2014developing}'s weighted lexicon of terms that uses the dataset of 75,394 Facebook users who shared their status, age and gender. The predictive power of this lexica was evaluated on Twitter, blog, and Facebook, showing promising results \cite{sap2014developing}. Utilizing these two weighted lexicon of terms, we are predicting the demographic information (age or gender) of $user_i$  (denoted by $Demo_{i}$) using following equation:
{\centering
$Demo_{i} = \sum_{terms\epsilon lex} Weight_{lex}(term)*\frac{Freq(term,doc)_{i}}{WC(doc)_{i}}$
}

where $Weight_{lex}(term)$ is the lexicon weight of the term, and 
$Freq(term,doc)_i$ represents the frequency of the term in the user generated $doc_i$, and $WC(doc)_i$ measures total word count in $(doc)_i$. As our data is biased toward young people, we report age prediction performance for each age group separately (Table \ref{Age_prediction}). Moreover, to measure the average accuracy of this model, we build a balanced dataset (keeping all the users above 23 -416 users), and then randomly sampling the same number of users from the age ranges (11,19] and (19,23].  The average accuracy of this model is 0.63 for depressed users and 0.64 for control class. Table \ref{Gedner_predcition} illustrates the performance of gender prediction for each class. The average accuracy is 0.82 on $U_g$ ground-truth dataset.

\noindent
\textbf{Prediction with Visual Imagery:}
\noindent
Inspired by \cite{zhou2013extensive}'s approach for facial landmark localization, we use their pretrained CNN consisting of convolutional layers, including unshared and fully-connected layers, to predict gender and age from both the \textit{profile} and \textit{shared images}.
We evaluate the performance for gender and age prediction task on $U_g$ and $U_a$ respectively as shown in Table \ref{Age_prediction} and Table \ref{Gedner_predcition}. 

\noindent
\textbf{Demographic Prediction Analysis:}
\noindent
We delve deeper into the benefits and drawbacks of each data modality for demographic information prediction. This is crucial as the differences between language cues between age groups above age 35 tend to become smaller (see Figure \ref{LIWC-age}-A,B,C) and making the prediction harder for older people \cite{eckert2017age}. In this case,  the other data modality (e.g., visual content) can play integral role as a complementary source for age inference.
For gender prediction (see Table \ref{Gedner_predcition}), on average, the profile image-based predictor provides a more accurate prediction for both the depressed and control class (0.92 and 0.90) compared to content-based predictor (0.82).
For age prediction (see Table \ref{Age_prediction}), textual content-based predictor (on average 0.60) outperforms both of the visual-based predictors (on average profile:0.51, Media:0.53).  
\begin{table*}
\centering
\tiny
\setlength{\abovecaptionskip}{1pt plus 0pt minus 2pt}
\caption{Age Prediction Performance from Visual and Textual Content for Different Age Group(in Years Old)}
\label{Age_prediction}
\begin{tabular}{c|c|cccc|cccc|cccc}
\hline
\multirow{2}{*}{Group} & \multirow{2}{*}{Measure} & \multicolumn{4}{c|}{Text-based}               & \multicolumn{4}{c|}{\begin{tabular}[c]{@{}c@{}}Image-based\\ (Profile)\end{tabular}} & \multicolumn{4}{c}{\begin{tabular}[c]{@{}c@{}}Image-based\\ (Media)\end{tabular}} \\ \cline{3-14} 
                       &                          & (11,19{]} & (19,23{]} & (23,34{]} & (34,46{]} & (11,19{]}           & (19,23{]}           & (23,34{]}           & (34,46{]}          & (11,19{]}           & (19,23{]}          & (23,34{]}          & (34,46{]}          \\ \hline
                       & Sensitivity              & 0.23      & 0.38      & 0.65      & 0.33      & 0.29                & 0.29                & 0.22                & 1.0                & 0.11                & 0.1                & 0.19               & 0.22               \\
Depressed              & Specificity              & 0.95      & 0.53      & 0.69      & 0.96      & 0.92                & 0.92                & 0.57                & 0.80               & 0.96                & 0.94               & 0.72               & 0.58               \\
                       & ACC                      & 0.59      & 0.46      & 0.67      & 0.65      & 0.47                & 0.46                & 0.40                & 0.900              & 0.50                & 0.49               & 0.46               & 0.40               \\ \hline
                       & Sensitivity              & 0.14      & 0.31      & 0.62      & 0.69      & 0.12                & 0.1                 & 0.40                & 0.25               & 0.15                & 0.30               & 0.63               & 0.64               \\
Control                & Specificity              & 0.98      & 0.63      & 0.61      & 0.90      & 0.90                & 0.95                & 0.53                & 0.75               & 0.98                & 0.62               & 0.60               & 0.91               \\
                       & ACC                      & 0.56      & 0.47      & 0.62      & 0.80      & 0.49                & 0.48                & 0.47                & 0.51               & 0.56                & 0.46               & 0.62               & 0.77               \\ \hline
\end{tabular}
\vspace{-10pt}
\end{table*}
However, not every user provides facial identity on his account (see Table \ref{facial-Presence}). We studied facial presentation for each age-group to examine any association between age-group, facial presentation and depressive behavior (see Table \ref{facial_presence_age}). We can see youngsters in both depressed and control class are not likely to present their face on profile image. Less than 3\% of vulnerable users between 11-19 years  reveal their facial identity. 
Although content-based gender predictor was not as accurate as image-based one, it is adequate for population-level analysis.
%profile: depressed 0.48. , control 0.55
%Media depressed 0.46 control 0.60 Text depressed, 0.59 Text Control, 0.61
%\subsubsection{Prediction via Textual Content}
\begin{table*}
\centering
\tiny
\setlength{\abovecaptionskip}{1pt plus 0pt minus 2pt}
\caption{Facial Presentation Distribution for Different Age Group(in Years Old) in Profile and Media}
\label{facial_presence_age}
\begin{tabular}{ll|l|l|l|l|lllll}
\hline
                               & \multicolumn{5}{c|}{\% Users Faces\_Found\_ in\_Profile
                               } & \multicolumn{5}{l}{\% Users Faces\_Found\_ in\_Media}                                                                                       \\ \hline
\multicolumn{1}{l|}{}          & {[}11,19)  & {[}19,23) & {[}23,34) & {[}34,46) & {[}46,60) & \multicolumn{1}{l|}{{[}11,19)} & \multicolumn{1}{l|}{{[}19,23)} & \multicolumn{1}{l|}{{[}23,34)} & \multicolumn{1}{l|}{{[}34,46)} & {[}46,60) \\ \hline
\multicolumn{1}{l|}{Control}   & 4.55       & 9.58      & 13.84     & 17.85     & 21.42     & \multicolumn{1}{l|}{89.70}     & \multicolumn{1}{l|}{88.35}     & \multicolumn{1}{l|}{78.46}     & \multicolumn{1}{l|}{67.85}     & 78.57     \\
\multicolumn{1}{l|}{Depressed} & 2.71       & 5.88      & 10.52     & 8.33      & 14.28     & \multicolumn{1}{l|}{90.21}     & \multicolumn{1}{l|}{90.58}     & \multicolumn{1}{l|}{76.31}     & \multicolumn{1}{l|}{83.33}     & 85.71    
\end{tabular}
\vspace{-22pt}
\end{table*}
\begin{table*}
\centering
\begin{tiny}
\setlength{\abovecaptionskip}{1pt plus 0pt minus 2pt}
\setlength{\tabcolsep}{4.35pt}
\caption{Gender Prediction Performance through Visual and Textual Content}
\label{Gedner_predcition}
%p{9mm}
\begin{tabular}{c|l|l|c|c|c|c|c|c|c|c|c|c|c|c|c|c|c|c}
\hline
\multicolumn{3}{c|}{\multirow{3}{*}{\begin{tabular}[c]{@{}c@{}}Face found \\in \end{tabular}}} & \multicolumn{4}{c|}{Agreement}                                & \multicolumn{6}{c|}{Image-based Predictor}                                                                                                                                              & \multicolumn{6}{c}{Content-based Predictor}                                                                                                                                                      \\ \cline{4-19} 
\multicolumn{3}{c|}{}                                                                                       & \multicolumn{2}{c|}{Depressed} & \multicolumn{2}{c|}{Control} & \multicolumn{3}{c|}{Depressed}                                                            & \multicolumn{3}{c|}{Control}                                                                & \multicolumn{3}{c|}{Depressed}                                                                                               & \multicolumn{3}{c}{Control}                                       \\ \cline{4-19} 
\multicolumn{3}{c|}{}                                                                                       & \begin{tabular}[c]{@{}c@{}}Cohen's\\ kappa\end{tabular}               & pct.           & \begin{tabular}[c]{@{}c@{}}Cohen's\\ kappa\end{tabular}              & pct.          & Sens. & Spec. & ACC (95\% CI)                                                 & Sens. & Spec. & ACC (95\% CI)                                                   & Sens.           & Spec.           & ACC (95\% CI)                                                                & Sens.           & Spec.           & ACC (95\% CI)      \\ \hline
\multicolumn{3}{c|}{\begin{tabular}[c]{@{}c@{}}Profile\\ \\ \end{tabular}}        & 0.32***         & 73.9         & 0.31***        & 70.3        & 0.90        & 1.0         & \begin{tabular}[c]{@{}c@{}}0.92\\ (0.80, 0.98)\end{tabular}   & 0.91        & 0.87        & \begin{tabular}[c]{@{}c@{}}0.90\\ (0.81, 0.95)\end{tabular}     & \multirow{4}{*}{0.87} & \multirow{4}{*}{0.50} & \multirow{4}{*}{\begin{tabular}[c]{@{}c@{}}0.82\\ (0.79, 0.85)\end{tabular}} & \multirow{4}{*}{0.86} & \multirow{4}{*}{0.76} & \multirow{4}{*}{\begin{tabular}[c]{@{}c@{}}0.82\\ (0.79, 0.85)\end{tabular}} \\ \cline{1-13}
\multicolumn{3}{c|}{\begin{tabular}[c]{@{}c@{}}Media\\ \\ \end{tabular}}         & 0.1*            & 53.4         & 0.09***        & 52.3        & 0.57        & 0.70        & \begin{tabular}[c]{@{}c@{}}0.584\\ (0.546, 0.62)\end{tabular} & 0.46        & 0.65        & \begin{tabular}[c]{@{}c@{}}0.51\\ (0.4634, 0.5595)\end{tabular} &                       &                       &                                                                              &                       &                       &                    \\ \hline
\end{tabular}
\end{tiny}
\vspace{-10pt}
\end{table*}

%\section{Predicting Depressive Behavior}
\section{Multi-modal Prediction Framework}
\begin{figure}[ht!]
    \centering
    \includegraphics[width=0.45\textwidth]{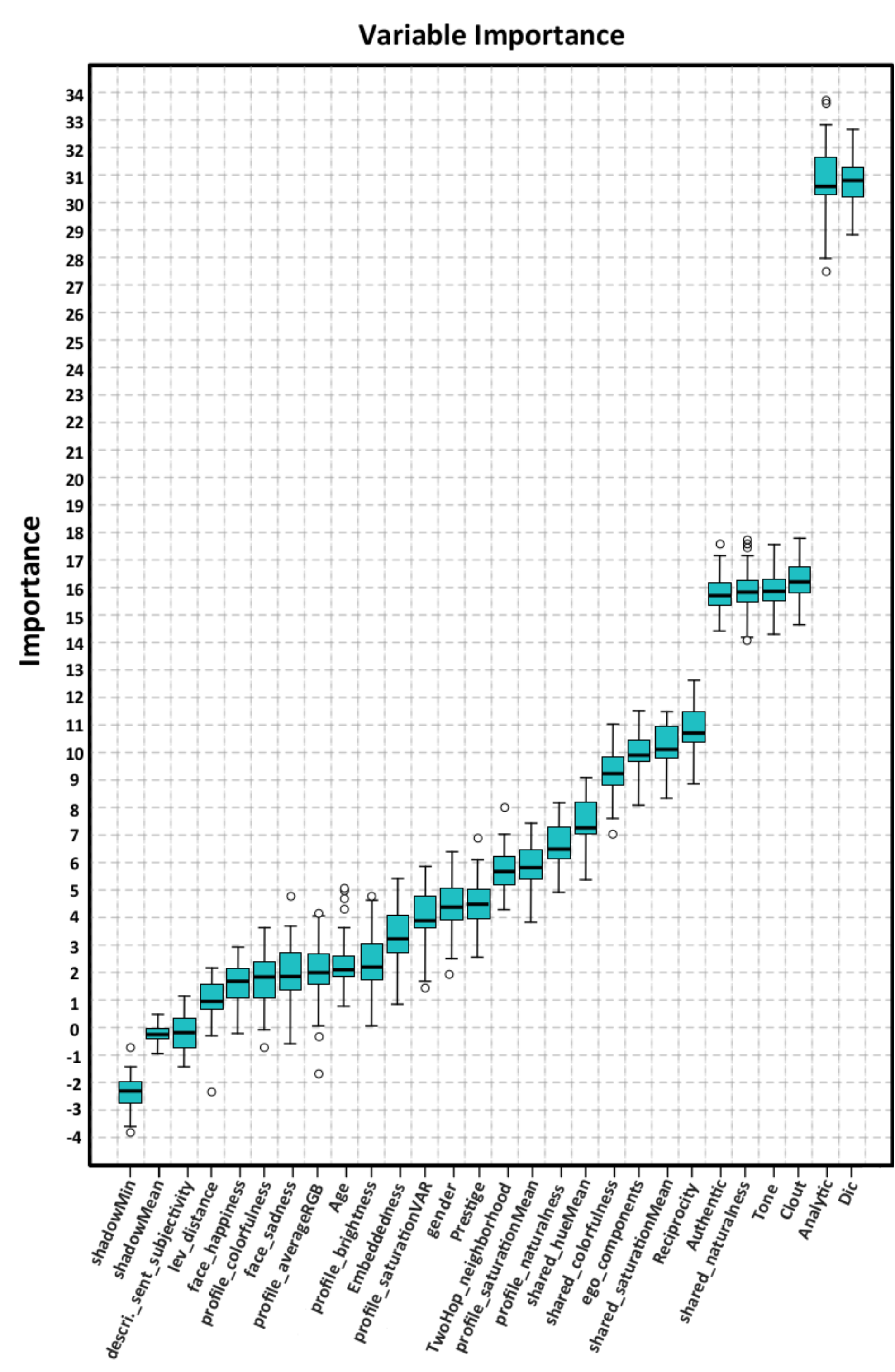}
    \setlength{\abovecaptionskip}{1pt plus 0pt minus 2pt}
    \caption{Ranking Features obtained from Different Modalities with an Ensemble Algorithm }
    \vspace{-6pt}
    \label{Feature_importance}
\end{figure}

We use the above findings for predicting depressive behavior. Our model exploits early fusion \cite{duong2017multimodal} technique in feature space and requires modeling each user $u$ in $U_t$ as vector concatenation of
individual modality features. As opposed to computationally expensive late fusion scheme where each modality requires a separate supervised modeling, this model reduces the learning effort and shows promising results \cite{snoek2005early}. To develop a generalizable model that avoids overfitting, we perform feature selection using statistical tests and \textit{all relevant} ensemble learning models. It adds randomness to the data by creating shuffled copies of all features (shadow feature), and then trains Random Forest classifier on the extended data. Iteratively, it checks whether the actual feature has a higher Z-score than its shadow feature (See Algorithm \ref{algo} and Figure \ref{Feature_importance}) \cite{kursa2010feature}.  

\begin{algorithm}
\Fn{Main}{
\For{each Feature $X_j \in X$}{
   \textit {$ShadowFeatures \leftarrow   RndPerm(X_j)$}}
 %\EndFor
   
 \textit {RndForrest($ShadowFeatures,X$)}\;
 \textit{ Calculate Imp $ (X_j, MaxImp(ShadowFeatures))$}\;
 \If {$Imp(X_j) > MaxImp(ShadowFeatures)$}{
       \textit{Generate next hypothesis}  , \Return $X_j$ 
 }
 \textit {Once all hypothesis generated}\;
 \textit{Perform} Statistical Test $H_0: H_i = E(H) vs H_1:H_i \neq E(H) \;
 H_i  \sim  N((0.5N)((\sqrt{0.25N})^2))$ \textit{//Binomial Distribution}\;
 \eIf {$H_i \gg  E(H)$}{\textit{Feature is important}}
 {\textit{Feature is important}}
 }
\vspace{-5pt}
\label{algo}
\caption{Ensemble Feature Selection}
\end{algorithm}

Next, we adopt an ensemble learning method that integrates the predictive power of multiple learners with two main advantages; its interpretability with respect to the contributions of each feature and its high predictive power. For prediction we have $y_i'= \sum_{t=1}^{m}f_t(u_i)$ where $f_t$ is a weak learner and $y_i'$ denotes the final prediction.

In particular, we optimize the loss function:
$L^{<t>}= \sum_{i=1}^{n}l(y_i,y_i^{{}'<t-1>}+f_t(u_i))+\varphi  (f_t)$
where $\varphi$ incorporates $L1$ and $L2$ regularization. In each iteration, the new $f_t(u_i)$ is obtained by fitting weak learner to the negative gradient of loss function. Particularly, by estimating the loss function with Taylor expansion \footnote{https://bit.ly/2Ga1c0w}:
$ L^{<t>}  \sim \sum_{i=1}^{n}l(y_i,y_i^{{}'<t-1>})+ (\frac{\partial l(y_i,y_i^{{}'<t-1>}}{\partial y_i^{{}'<t-1>}})f_t(u)+(\frac{\partial^2 l(y_i,y_i^{{}'<t-1>}}{\partial y_i^{{}'<t-1>^2}})f_t(u_i)^2\ $
where its first expression is constant, the second and the third expressions are first ($g_i$) and second order derivatives ($h_i$) of the loss.
$$L^{<t>}  = \sum_{i=1}^{n}(g_if_t(u_i)+h_if_t(u_i))+ \varphi (f_t)$$

\begin{figure}[ht!]
    \centering
    \includegraphics[width=0.45\textwidth]{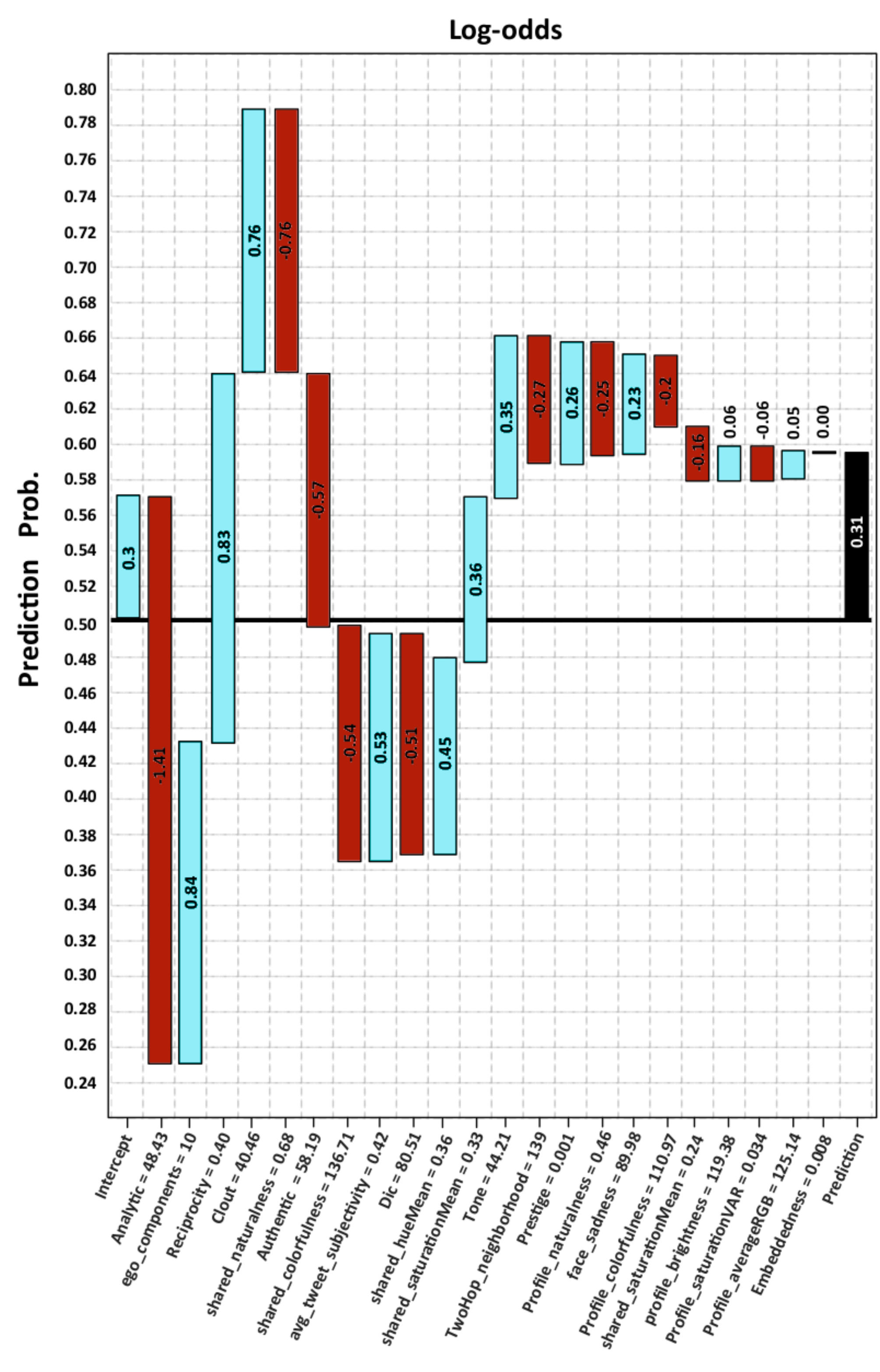}
    \setlength{\abovecaptionskip}{2pt plus 0pt minus 3pt}
    \caption{The explanation of the log-odds prediction of outcome (0.31) for a sample user (y-axis shows the outcome probability (depressed or control), the bar labels indicate the log-odds impact of each feature)}
    \vspace{-6pt}
    \label{Feature_contrib}
\end{figure}

For exploring the weak learners, assume $f_t$ has k leaf nodes, $I_j$ be subset of users from $U_t$ belongs to the node $j$, and $w_j$ denotes the prediction for node $j$. Then, for each user $i$ belonging to $I_j$, $f_t(u_i)=w_j$ and $\varphi (f_t) =  1/2 \lambda \sum_{j=1}^{k}W_j^2 + \gamma k $
$$ L^{<t>}  = \sum_{j=1}^{k}[(\sum_{i \epsilon I_j}g_i)w_j+1/2(\sum_{i \epsilon I_j}h_i+\lambda)w_j^2)]+\gamma k$$
Next, for each leaf node $j$, deriving w.r.t $w_j$:
$$w_j = \frac{-\sum_{i \epsilon I_i}g_i}{\sum_{i \epsilon I_j}h_i+\lambda}$$ and by substituting weights:
$$L^{<t>}= -1/2\sum_{j=1}^{k} \frac{(\sum_{i \epsilon I_j}g_i)^2}{\sum_{i \epsilon I_j}h_i+\lambda} + \gamma k$$
which represents the loss for fixed weak learners with $k$ nodes. The trees are built sequentially such that each subsequent tree aims to reduce the errors of its predecessor tree. Although, the weak learners have high bias, the ensemble model produces a strong learner that effectively integrate the weak learners by reducing bias and variance (the ultimate goal of supervised models) \cite{chen2016xgboost}. Table \ref{performance_table} illustrates our multimodal framework outperform the baselines for identifying depressed users in terms of average specificity, sensitivity, F-Measure, and accuracy in 10-fold cross-validation setting on $U_t$ dataset. Figure \ref{Feature_contrib} shows how the likelihood of being classified into the depressed class varies with each feature 
addition to the model for a sample user in the dataset.
The prediction bar (the black bar) shows that the log-odds of prediction is 0.31,  that is,  the likelihood of this person being a depressed user is  57\% (1 / (1 + exp(-0.3))). The figure also sheds light on the impact of each contributing feature. The waterfall charts represent how the probability of being depressed changes with the addition of each feature variable. For instance, the "Analytic thinking" of this user is considered high 48.43 (Median:36.95, Mean: 40.18) and this decreases the chance of this person being classified into the depressed group by the log-odds of -1.41. Depressed users have significantly lower \textit{"Analytic thinking"} score compared to control class. Moreover, the 40.46  \textit{"Clout"} score is a low value (Median: 62.22, Mean: 57.17) and it decreases the chance of being classified as depressed. 
With respect to the visual features, for instance, the  mean and the median of 'shared\_colorfulness' is 112.03 and 113 respectively. The value of 136.71 would be high; thus, it decreases the chance of being depressed for this specific user by log-odds of -0.54. Moreover, the 'profile\_naturalness' of 0.46 is considered high compared to 0.36 as the mean for the depressed class which justifies pull down of the log-odds by $-0.25$. For network features, for instance, 'two\_hop\_neighborhood' for depressed users (Mean : 84) are less than that of control users (Mean: 154), and is reflected in pulling down the log-odds by -0.27.
%For network features, our results are consistent with \cite{de2013social} where, for instance, \textit{"two\_hop\_neighborhood"} for depressed users (Mean:84) are less than that of control users (Mean:154), and is reflected in pulling down the log-odds by -0.27.  Similarly, for \textit{Embeddedness}, both the mean and median for depressed users are larger; however, for this specific instance, the difference is trivial, such that the log-odds remains constant. 

\noindent
\textbf{Baselines:}
\noindent
To test the efficacy of our multi-modal framework for detecting depressed users, we compare it against existing content, content-network, and image-based models (based on the aforementioned general image feature, facial presence, and facial expressions.) 
\begin{table*}
\tiny
\setlength{\abovecaptionskip}{1pt plus 0pt minus 2pt}
\caption{Model's Performance for Depressed User Identification from Twitter using different data modalities}
\label{performance_table}
\begin{tabular}{p{4mm}|p{12mm}|l|llllp{2.5mm}p{2.5mm}H@{\hspace*{-\tabcolsep}}|p{17mm}|p{2.5mm}|p{3mm}|p{2.5mm}|p{2.5mm}}
%p{1.5mm}|p{8.2mm}
\hline
\multirow{2}{*}{\begin{tabular}[c]{@{}c@{}}Model\\ \#\end{tabular}} & \multicolumn{1}{c|}{\multirow{2}{*}{Data Source}} & \multirow{2}{*}{Ref}                               & \multicolumn{1}{l|}{\multirow{2}{*}{Year}} & \multicolumn{6}{c|}{Features}                                                                                                                                            & \multirow{2}{*}{Model}     & \multirow{2}{*}{Spec.} & \multirow{2}{*}{Sens.} & \multirow{2}{*}{F-1} & \multirow{2}{*}{Acc.} \\ \cline{5-10}
                                                                    & \multicolumn{1}{c|}{}                             &                                                    & \multicolumn{1}{l|}{}                      & \multicolumn{1}{l|}{N-grams} & \multicolumn{1}{l|}{LIWC} & \multicolumn{1}{l|}{Sentiment} & \multicolumn{1}{l|}{Topics} & \multicolumn{1}{l|}{Metadata} & others         &                            &                              &                              &                      &                           \\ \hline
I                                                                   & \multicolumn{1}{c|}{}                             & \cite{nadeem2016identifying}      & \multicolumn{1}{l|}{2016}                  & \multicolumn{1}{l|}{X}       & \multicolumn{1}{l|}{}     & \multicolumn{1}{l|}{}          & \multicolumn{1}{l|}{}       & \multicolumn{1}{l|}{}         &                & NB                         & 0.69                         & 0.70                         & 0.69                 & 0.70                      \\
II                                                                  &                                                   & \cite{coppersmith2016exploratory} & \multicolumn{1}{l|}{2016}                  & \multicolumn{1}{l|}{X}       & \multicolumn{1}{l|}{}     & \multicolumn{1}{l|}{X}         & \multicolumn{1}{l|}{}       & \multicolumn{1}{l|}{User Acti.}        & User Acti.  & Not Reported*              & 0.73                         & 0.74                         & 0.73                 & 0.74                      \\
III                                                                 &                                                   & \cite{coppersmith2014quantifying} & \multicolumn{1}{l|}{2015}                  & \multicolumn{1}{l|}{X}       & \multicolumn{1}{l|}{X}    & \multicolumn{1}{l|}{X}         & \multicolumn{1}{l|}{}       & \multicolumn{1}{l|}{User Acti.}         & User Acti.  & Log-linear       & 0.83                         & 0.80                         & 0.81                 & 0.82                      \\
IV                                                                  &                                                   & \cite{preoctiuc2015role}          & \multicolumn{1}{l|}{2015}                  & \multicolumn{1}{l|}{X}       & \multicolumn{1}{l|}{X}    & \multicolumn{1}{l|}{X}         & \multicolumn{1}{l|}{X}      & \multicolumn{1}{l|}{}         & Gender     & LR                         & 0.84                         & 0.83                         & 0.84                 & 0.84                      \\
V                                                                   & Content                                           & \cite{tsugawa2015recognizing}     & \multicolumn{1}{l|}{2015}                  & \multicolumn{1}{l|}{X}       & \multicolumn{1}{l|}{X}    & \multicolumn{1}{l|}{X}         & \multicolumn{1}{l|}{X}      & \multicolumn{1}{l|}{User Acti.}        & User Acti.  & SVM                        & 0.86                         & 0.84                         & 0.85                 & 0.85                      \\
VI                                                                &                                                   & N/A                                       & \multicolumn{1}{l|}{N/A}                   & \multicolumn{1}{l|}{X}       & \multicolumn{1}{l|}{}     & \multicolumn{1}{l|}{}          & \multicolumn{1}{l|}{}       & \multicolumn{1}{l|}{}         &                & SVM(Pre. embed.) & 0.72                         & 0.72                         & 0.72                 & 0.72                      \\
VII                                                                  &                                                   & N/A                                        & \multicolumn{1}{l|}{N/A}                   & \multicolumn{1}{l|}{X}       & \multicolumn{1}{l|}{}     & \multicolumn{1}{l|}{}          & \multicolumn{1}{l|}{}       & \multicolumn{1}{l|}{}         &                & SVM(Train w2vec)       & 0.70                         & 0.70                         & 0.70                 & 0.70                      \\ \hline
VIII                                                                   & Cont., Net.                                   & \cite{de2013predicting}           & \multicolumn{1}{l|}{2013}                  & \multicolumn{1}{l|}{X}       & \multicolumn{1}{l|}{X}    & \multicolumn{1}{l|}{X}         & \multicolumn{1}{l|}{}       & \multicolumn{1}{l|}{}         &  Network & SVM, PCA                   & 0.84                             &0.80                              &  0.83                    &   0.85                        \\ \hline
IX                                                                  & \multicolumn{1}{c|}{\multirow{3}{*}{Image}}       & N/A                                        & \multicolumn{1}{c|}{N/A}                   & \multicolumn{6}{c|}{\multirow{3}{*}{N/A}}                                                                                                                                & LR                         & 0.68                         & 0.67                        & 0.67                 & 0.68                      \\ \cline{1-1}
X                                                                 & \multicolumn{1}{c|}{}                             & N/A                                        & \multicolumn{1}{l|}{N/A}                   & \multicolumn{6}{c|}{}                                                                                                                                                    & SVM                        & 0.69                         & 0.67                         & 0.67                 & 0.69                      \\ \cline{1-1}
XI                                                                & \multicolumn{1}{c|}{}                             & N/A                                       & \multicolumn{1}{l|}{N/A}                   & \multicolumn{6}{c|}{}                                                                                                                                                    & RF                         & 0.72                         & 0.70                         & 0.69                 & 0.71                      \\ \hline
\textbf{Ours}                                                                   & Cont.,Image,Net.                             &    N/A                                                & X                                          & X                            & X                         & X                              & X                           & X                             & \makecell{Demog.\\ User Acti.}               & N/A                    & \textbf{0.87}                         & \textbf{0.92}                         & \textbf{0.90}                 & \textbf{0.90}                      \\ \hline

\end{tabular}
\vspace{-20pt}
\end{table*}

\noindent
\textbf{Content-based models:}
\noindent
See table \ref{performance_table} for the performance of our prediction framework against the state-of-the-art methods for predicting depressive behavior employing the same feature sets and hyperparameter settings (see \textbf{Models I-V}.) Besides, several prior efforts demonstrate that word embedding models can reliably enhance short text classification \cite{wang2015semantic}, \textbf{Model VI} employs pre-trained word embeddings trained over 400 million tweets \footnote{\url{https://bit.ly/2sPR3OQ}} while representing a user with retrieving word vectors for all the words a user employed in tweets/profile description. We aggregate these word vectors through their means and feeding it as input to SVM classifier with a linear kernel. In \textbf{Model VII}, we employ \cite{yazdavar2017semi}'s dataset of 45000 self-reported depressed users and train Skip-
gram model with negative sampling to learn word representations. We chose this model as it generates robust word embeddings even when words are sparse in the training corpus \cite{mikolov2013distributed}. We set dimensionality to 300 and negative sampling rate to 10 sample words, which shows promising results with medium-sized datasets \cite{mikolov2013distributed}. Besides, we observed many vulnerable users chose specific account names, such as "Suicidal\_Thoughxxx," and "younganxietyyxxx," which are good indicators of their depressive behavior.  We use Levenshtein distance \footnote{\url{https://bit.ly/1JtgTWJ}} between depression indicative terms in \cite{yazdavar2017semi}'s depression lexicon and the screen name to capture their degree of semantic similarity.

\noindent
\textbf{Image-based models:}
\noindent
We employ the aforementioned visual content features including facial presence, aesthetic features, and facial expression for depression prediction. We use three different models: Logistic Regression (\textbf{Model IX}), SVM (\textbf{Model X}), and Random Forrest (\textbf{Model XI}). The poor performance of image-based models suggests relying on a unique modality would not be sufficient for building a robust model given the complexity and the abstruse nature of prediction task.

\noindent
\textbf{Network-based models:}
\noindent
Network-based features imply users' desire to socialize and connect with others. There is a notable difference between number of friends and followers, favorites and status count for depressed and control users (see Table \ref{t-test}.) Besides, for building baseline \textbf{Model VIII}, we obtained egocentric network measures for each user based on the network formed using @-replies interactions among them. The egocentric social graph of a user u is an undirected graph of nodes in u's two-hop neighborhood in our ${U_a}$ dataset, where the edge between nodes u and v implies that there has been at least one @-reply exchange. Network-based features including \textit{Reciprocity}, \textit{Prestige Ratio}, \textit{Graph Density}, \textit{Clustering Coefficient}, \textit{Embeddedness}, \textit{Ego components} and \textit{Size of two-hop neighborhood} were extracted from user's network \cite{de2013predicting} for reliable capturing of user context for depression prediction. 

%High values for the three metrics -clustering coefficient, embeddedness and the number of ego networks - indicate that the depressed users tend to build a close network of trusted people to share their mental health issues. 
%High values for the three metrics -- clustering coefficient, embeddedness and the number of ego networks --- indicate that the depressed users tend to build a close network of trusted people to share their mental health issues. For both graph density and size of the two-hop neighborhood,  lower the value,  less the interactions.

%\subsection{Multi-modal Prediction Framework}

\section{Conclusion }

We presented an in-depth analysis of visual and contextual content of likely depressed profiles on Twitter. We employed them for demographic (age and gender) inference process. We developed multimodal framework, employing statistical techniques for fusing heterogeneous sets of features obtained by processing visual, textual and user interactions. Conducting extensive set of experiments, we assessed the predictive power of our multimodal framework while comparing it against state-of-the-art approaches for depressed user identification on Twitter. The empirical evaluation shows that our multimodel framework is superior to them and it improved the average F1-Score by 5 percent. Effectively, visual cues gleaned from content and profile shared on social media can further augment inferences from textual content for reliable determination of depression indicators and diagnosis. 
\newpage
\bibliography{ref}
\bibliographystyle{acl_natbib}

\end{document}